\documentclass[preprint,showpacs,aps]{revtex4-1}
\usepackage{soul}
\usepackage{amsmath}
\usepackage{graphicx}
\usepackage{latexsym}
\usepackage{verbatim}
\newcommand{\ba}{\begin{eqnarray}}
\newcommand{\ea}{\end{eqnarray}}

\usepackage{hyperref}
\newcommand{\nn}{\nonumber}

\newcommand{\bs}{\boldsymbol} 
\usepackage{color}
\begin{document}
\title{\bf\large{Quark Wigner Distributions Using Light-Front Wave Functions}}
\author{\bf Jai More$^{a}$, Asmita Mukherjee$^{a}$ and Sreeraj Nair$^{b}$}

\affiliation{$^{a}$  Department of Physics,
Indian Institute of Technology Bombay,\\ Powai, Mumbai 400076,
India.\\
$^{b}$ Department of Physics, Indian Institute of Science Education and Research Bhopal\\
Bhopal Bypass Road, Bhauri, Bhopal 462066 India}

\date{\today}

\begin{abstract}
The quasiprobabilistic Wigner distributions 
 are the quantum mechanical analog of the classical phase space distributions. 
We investigate quark Wigner distributions for a quark state dressed with a gluon, 
which can be thought of as a simple composite and relativistic  spin-1/2 state with a gluonic degree of freedom.
We calculate various polarization configurations, namely  unpolarized, longitudinally polarized, and 
transversely polarized quark, and the target state using   light-front wave functions in this model. 
At the leading twist, one can define 16 quark Wigner distributions, however, we obtain only 8 independent nonzero Wigner distributions in our model. We compare our results with other model calculations for the proton.
\end{abstract}
\maketitle
\section{Introduction}
The holy grail in the field of hadron physics is to come up with a better understanding 
of the structure of hadrons in terms of quarks and gluons.
Although the theory of quantum chromodynamics (QCD) explains most aspects of strong interaction, 
its nonperturbative nature makes it difficult to do  {\it ab initio} calculations.
In particular, calculating the spin correlations and momentum distribution of the fundamental building blocks 
inside the parent hadron has proven to be a challenge.
In overcoming this challenge, generalized parton distributions (GPDs) \cite{Mueller98,Goeke01,Diehl03,Ji04,Belitsky05,Boffi07} and transverse momentum dependent
parton distributions (TMDs) \cite{Collins81uk,Collins81uw,Mulders95,Sivers89,Kotzinian94,Boer97} have played an important role. 
GPDs, which were introduced experimentally in the context of deeply virtual Compton scattering (DVCS), are defined using 
off-forward matrix elements \cite{Ji96,Brodsky06,Radyushkin97}.
GPDs contain simultaneous information about the longitudinal momentum and transverse position  distribution of the partons.
GPDs have triggered interest mainly due to two reasons. Firstly their impact parameter representation \cite{Burkardt00,Diehl02,DC05} gives a probabilistic interpretation of finding a quark with longitudinal momentum fraction $x$ at a distance $b$ from the center of the target and secondly they contain information on the elusive
orbital angular momentum of the partons \cite{Ji97,Hagler03,Kanazawa14,Rajan16}. 
TMDs are accessed experimentally via semi-inclusive deep inelastic scattering (SIDIS) and the Drell-Yan process. 
In addition to the longitudinal momentum fraction, TMDs encode information
about the momentum distribution in the transverse direction. 
TMDs have been proven to be vital tools for doing three-dimensional nucleon tomography \cite{Radici14} in momentum space and they also provide correlation between the spin and orbital angular momentum of quarks.

A most general correlator that contains the maximum amount of information about the constituents
inside a hadron is the fully unintegrated, off-diagonal quark-quark correlator called the generalized parton correlation functions (GPCFs) introduced in \cite{Meissner09,Lorce13}. 
Integrating out the quark light-cone energy from GPCFs gives us the generalized transverse momentum dependent parton distributions 
(GTMDs)\cite{Liuti13,Burkardt15}. Both the GPDs and TMDs are related to GTMDs under appropriate limiting conditions and hence GTMDs can also be called  their ``mother distributions''. 
Wigner distributions \cite{Wigner32} can be thought of as the quantum analog of the classical phase space distributions
and they are related to the GTMDs via a Fourier transform. However, being a quantum distribution, it is constrained by the uncertainty principle and as a result, Wigner distributions are not positive definite over
the entire phase space. So Wigner distributions do not have a probabilistic interpretation. Nevertheless,
they are  very useful tools in understanding quark/gluon spin and angular momentum correlations inside the target nucleon, and certain model-based relations, and under certain
conditions, it is possible to have a semiclassical interpretation \cite{Lorce11}.
Moreover, Wigner distributions have previously been studied in various fields such
as quantum molecular dynamics, quantum information, image processing etc. \cite{Balazs83,Hillery83,Lee95} and in fact, there are experiments in which it is measured \cite{Vogel89,Smithey93,Breitenbach97,Banaszek99}.
In QCD, the Wigner distributions were first looked into using the nonrelativistic 
approximation in Refs.\cite{Ji03,Belitsky03} where
the distribution was studied as a six-dimensional function. The six-dimensional space consists of three position and three momentum coordinates. 
Then in Ref. \cite{Lorce11}, the authors defined a five-dimensional   
Wigner distribution consistent with relativity using the light-cone framework. The five-dimensional space consists of two transverse position and three momentum coordinates.
Various phenomenological models like the light-cone constituent quark model 
\cite{Lorce12}, chiral quark soliton model \cite{Pasquini11}, light-front dressed quark model \cite{Asmita14,Asmita15}, light-cone spectator model \cite{Liu14, Liu15} and  diquark model \cite{Miller14, Muller14} have been used to study
Wigner distributions. A complete multipole analysis of the quark Wigner distributions 
including transverse polarization was recently studied \cite{Lorce16}.

In this work, we study the Wigner distribution of quarks using the light-front Hamiltonian 
gauge-fixed formulation \cite{Hari99}. The Hamiltonian approach in front form is advantageous 
compared to the conventional equal-time form mainly because of the absence of the square-root 
operator in the bound state eigenvalue equation and due to the triviality of the QCD vacuum 
structure.  Instead of a proton state, we take a simple composite spin-$1/2$ state, namely 
a quark dressed at one loop with a gluon.  Like the proton state, the dressed quark state can also 
be expanded in multiparticle occupation number Fock 
states and because of the trivial vacuum such an expansion gives a complete basis for 
diagonalizing the full theory \cite{Brodsky98}. The advantage is that unlike the proton 
 light-front wave functions (LFWFs),
 the two-particle LFWFs of the dressed quark state can be calculated analytically in perturbation theory,
 and thus this can be thought of as a field theory based perturbative model having a gluonic degree of freedom. 

In our previous work \cite{Asmita14} we
studied the three independent quark Wigner distributions for unpolarized and longitudinal 
polarization of quark and the target state. 
Now, we present the complete study involving unpolarized, longitudinal, and transverse 
polarization combinations of the target state as well as the quark, which results in five additional 
independent distributions.
For our numerical calculations, we adopt a better integration strategy called the Levin method \cite{Levin82,Levin96,Levin97}, which suits our oscillatory integrands. 
Thus, the numerical calculations are performed using an improved method over previous distributions and 
we also present a calculation of the new distributions. The preliminary work conferred in \cite{Jai15} is now elaborated in this paper.

The paper is organized in the following manner. In Sec. \ref{WD}, we start 
by giving the field theory definition of the quark Wigner distribution. 
We use the truncated Fock expansion for the dressed quark 
and express all the Wigner distributions in terms of overlaps 
of LFWFs. One can write 16 distributions at the 
leading twist after taking into account various polarization combinations; 
however, we obtain only eight independent Wigner distributions in this model that can be studied. 
In Sec. \ref{numerical}, we explain about the numerical strategy used for studying the 
Wigner distribution, which is a very important part of this work. Then in Sec. \ref{result}, we 
apply the numerical technique used to study the eight 
distributions in transverse momentum space, transverse position space and mixed space. 
Finally, we end by giving our conclusions in Sec. \ref{conclusion}. 

\section{Quark Wigner Distributions in dressed quark model}\label{WD}

The Wigner distribution of quarks can be defined as the Fourier transform of the quark-quark correlators defining the GTMDs 
\cite{Meissner09, Lorce11}
\ba
\rho^{[\Gamma]} ({\bs b}_{\perp},{\bs k}_{\perp},x,s,s') = \int \frac{d^2 \Delta_{\perp}}{(2\pi)^2} e^{-i {\bs \Delta}_{\perp}.{\bs b}_{\perp}}
W_{s\, s'}^{[\Gamma]} ({\bs \Delta}_{\perp},{\bs k}_{\perp},x) 
\ea
where ${\bs b_\perp}$ is the impact parameter space  conjugate to ${\bs \Delta}_\perp$, which is the  momentum transfer of a dressed 
quark in the transverse direction. 
GTMDs are defined through the quark-quark correlator $W^{[\Gamma]}$ at a fixed light-front time as 
\ba
W_{s\,s'}^{[\Gamma]} ({\bs \Delta}_{\perp},{\bs k}_{\perp},x)
&=&\int \!\!\frac{dz^{-}d^{2} {\bs z}_{\perp}}{2(2\pi)^3}e^{i k.z}
 \Big{\langle}p^{+},\frac{{\bs \Delta}_{\perp}}{2},s' \Big{|}
\overline{\psi}(-\frac{z}{2})\Omega \Gamma \psi(\frac{z}{2}) \Big{|}
p^{+},-\frac{{\bs \Delta}_{\perp}}{2},s\Big{\rangle }  \Big{|}_{z^{+}=0}
\ea
The initial and final dressed quark states are defined in the symmetric frame, with the average four-momentum of the dressed quark 
as $P={1 \over 2} (p'+p)$, the four-momentum transfer $\Delta=p'-p$, and $\Delta^+=0$. The longitudinal momentum is $p^+$, the 
transverse momentum transfer is  ${\bs \Delta}_\perp$ and $s$($s'$) is the helicity of the initial (final) target state. The average four momentum of the quark is $k$, 
with $k^+=x P^+$, where $x$ is the longitudinal momentum fraction of the parton. $\Omega$ is the gauge link and is chosen to be unity.

The state of a dressed quark with momentum $p$ and fixed helicity $s$ can be written in terms of light-front wave functions (LFWFs) as the perturbative expansion of the Fock state
\ba
  \Big{| }p^{+}, {\bs p}_{\perp}, s \Big{\rangle} &=& \Phi^{s}(p) b^{\dagger}_{s}(p) | 0 \rangle +
 \sum_{s_1 s_2} \int \frac{dp_1^{+}d^{2}p_1^{\perp}}{ \sqrt{16 \pi^3 p_1^{+}}}
 \int \frac{dp_2^{+}d^{2}p_2^{\perp}}{ \sqrt{16 \pi^3 p_2^{+}}} \sqrt{16 \pi^3 p^{+}}
 \delta^3(p-p_1-p_2) \nn \\[1.5ex] 
 &&\times\Phi^{s}_{s_1 s_2}(p;p_1,p_2) b^{\dagger}_{s_1}(p_1) 
 a^{\dagger}_{s_2}(p_2)  | 0 \rangle 
 \ea
where $\Phi^{s}(p)$ is the single quark state and $\Phi^{s}_{s_1 s_2}(p;p_1,p_2)$ is the quark gluon state LFWF. $\Phi^{s}(p)$ is the wave function normalization constant of the quark. $\Phi^{s}_{s_1 s_2}(p;p_1,p_2)$ 
gives the probability amplitude to find a bare quark (gluon) with momentum $p_1 (p_2)$ and helicity $s_1 (s_2)$ inside the dressed quark.
Using the Jacobi momenta
\ba k_{i}^{+} = x_{i}P^+ ~\text{and}~~~ {\bs k}_{i}^{\perp} ={\bs q}_{i}^{\perp} +  x_{i}{\bs P}^{\perp} \ea
so that 
\ba\sum_i x_i=1,~~~~~\sum_i {\bs q}_{i\perp}=0\ea 
The two-particle LFWF can be written in terms of the boost-invariant LFWF as
\ba\sqrt{P^+}\Phi(p; p_1, p_2) = \Psi(x_{i},{\bs q}_{i}^{\perp})\ea
The two-particle LFWF is given by \cite{Hari99} 
\ba
\Psi^{sa}_{s_1 s_2}(x,{\bs q}^{\perp})&=& 
\frac{1}{\Big[m^2 - \frac{m^2 + ({\bs q}^{\perp})^2 }{x} - \frac{({\bs q}^{\perp})^2}{1-x} \Big]}
\frac{g}{\sqrt{2(2\pi)^3}} T^a \chi^{\dagger}_{s_1} \frac{1}{\sqrt{1-x}}\nn\\
&\times&
 \Big[ -\frac{2{\bs q}^{\perp}}{1-x}   -  \frac{({\bs \sigma}^{\perp}.{\bs q}^{\perp}){\bs \sigma}^{\perp}}{x}
+\frac{im~{\bs \sigma}^{\perp}(1-x)}{x}\Big]
\chi_s ({\bs \epsilon}^{\perp}_{s_2})^{*}
\ea
Using two-component formalism \cite{Zhang93}, $\chi$, $T^a$, $m$ and ${\bs \epsilon}^{\perp}_{s2}$ are the two-component spinor, color 
SU(3) matrices, mass of the quark, and polarization 
vector of the gluons, respectively. 
At leading twist, one obtains only four Dirac operators $\Gamma=\{\gamma^+, \gamma^+\gamma^5,  i \sigma^{+1}\gamma^{5}, 
i \sigma^{+2}\gamma^{5}\}$, which corresponds to  Wigner distributions for unpolarized, 
longitudinally polarized, and transversely polarized dressed quark.  
So the quark-quark correlator using two-particle LFWFs for different polarizations at twist-2 is given by 
\ba\label{U}
W_{s\, s'}^{[\gamma^+]} ({\bs \Delta}_{\perp},{\bs k}_{\perp},x) & =&\sum_{\lambda_1',\lambda_{1},\lambda_2} 
\Psi^{*s'}_{\lambda_{1}' \lambda_2}(x,{\bs q'}^{\perp}) \chi_{\lambda_1'}^{\dagger} \chi_{\lambda_1}
\Psi^{s}_{\lambda_1 \lambda_2}(x,{\bs q}^{\perp})
\ea
\ba\label{L}
W_{s\, s'}^{[\gamma^+\gamma^{5}]} ({\bs \Delta}_{\perp},{\bs k}_{\perp},x) & =&\sum_{\lambda_1',\lambda_{1},\lambda_2} 
\Psi^{*s'}_{\lambda_{1}' \lambda_2}(x,{\bs q'}^{\perp}) \chi_{\lambda_1'}^{\dagger}\sigma_3  \chi_{\lambda_1}
\Psi^{s}_{\lambda_1 \lambda_2}(x,{\bs q}^{\perp})
\ea 
\ba\label{T}
W_{s\, s'}^{[i\sigma^{+j}\gamma^{5}]} ({\bs \Delta}_{\perp},{\bs k}_{\perp},x) & =&\sum_{\lambda_1',\lambda_{1},\lambda_2}  
\Psi^{*s'}_{\lambda_{1}' \lambda_2}(x,{\bs q'}^{\perp}) \chi_{\lambda_1'}^{\dagger}\sigma_j  \chi_{\lambda_1}
\Psi^{s}_{\lambda_1 \lambda_2}(x,{\bs q}^{\perp})
\ea
where $\sigma_i$ are the three Pauli matrices.
Equations (\ref{U}), (\ref{L}), and (\ref{T}) give unpolarized,
longitudinally polarized and transversely polarized GTMDs in terms of LFWFs.
For various combinations of unpolarized (U), longitudinally polarized (L),
and transversely polarized (T) target and quark states, the quark-quark correlators can be 
parametrized into 16 Wigner distributions \cite{Liu15} at leading twist. 
We denote Wigner distributions by $\rho_{\lambda,\lambda'}$, where $\lambda$ and $\lambda'$ 
represent the polarization of the target state and quark, respectively. The 16 possible leading twist
quark Wigner distributions are defined as follows.
\subsection{Unpolarized target and different quark polarization}

\noindent
The unpolarized Wigner distribution 
\ba
\rho_{UU}({\bs b}_\perp, {\bs k}_\perp,x)&=&\frac1{2}\Big[\rho^{[\gamma^+]}({\bs b}_\perp, {\bs k}_\perp,x,\hat{\bs e}_z)+\rho^{[\gamma^+]}({\bs b}_\perp, {\bs k}_\perp,x,-\hat{\bs e}_z)\Big]
\label{rhouu} 
\ea
The unpolarized-longitudinally polarized Wigner distribution
\ba
\rho_{UL}({\bs b}_\perp, {\bs k}_\perp,x)&=&\frac1{2}\Big[\rho^{[\gamma^+\gamma^5]}({\bs b}_\perp, {\bs k}_\perp,x,\hat{\bs e}_z)+\rho^{[\gamma^+\gamma^5]}({\bs b}_\perp, {\bs k}_\perp,x,-\hat{\bs e}_z)\Big]
\ea
The unpolarized-transversely polarized Wigner distribution
\ba
\rho^j_{UT}({\bs b}_\perp, {\bs k}_\perp,x)&=&\frac1{2}\Big[\rho^{[i \sigma^{+j}\gamma^5]}({\bs b}_\perp, {\bs k}_\perp,x,\hat{\bs e}_z)+\rho^{[i \sigma^{+j}\gamma^5]}({\bs b}_\perp, {\bs k}_\perp,x,-\hat{\bs e}_z)\Big]
\ea

\subsection{Longitudinal polarized target and different quark polarization}

\noindent
The longitudinal-unpolarized Wigner distribution
\ba
\rho_{LU}({\bs b}_\perp, {\bs k}_\perp,x)&=&\frac1{2}\Big[\rho^{[\gamma^+]}({\bs b}_\perp, {\bs k}_\perp,x,\hat{\bs e}_z)-\rho^{[\gamma^+]}({\bs b}_\perp, {\bs k}_\perp,x,-\hat{\bs e}_z)\Big]
\ea
The longitudinal Wigner distribution
\ba\rho_{LL}({\bs b}_\perp, {\bs k}_\perp,x)&=&\frac1{2}\Big[\rho^{[\gamma^+\gamma^5]}({\bs b}_\perp, {\bs k}_\perp,x,\hat{\bs e}_z)-\rho^{[\gamma^+\gamma^5]}({\bs b}_\perp, {\bs k}_\perp,x,-\hat{\bs e}_z)\Big]
\label{rholl}
\ea
The longitudinal-transversely-polarized Wigner distribution
\ba
\rho^j_{LT}({\bs b}_\perp, {\bs k}_\perp,x)&=&\frac1{2}\Big[\rho^{[i \sigma^{+j}\gamma^5]}({\bs b}_\perp, {\bs k}_\perp,x,\hat{\bs e}_z)-\rho^{[i \sigma^{+j}\gamma^5]}({\bs b}_\perp, {\bs k}_\perp,x,-\hat{\bs e}_z)\Big]
\ea

\subsection{Transversely polarized target and different quark polarization}

\noindent
The transverse-unpolarized  Wigner distribution
\ba
\rho^i_{TU}({\bs b}_\perp, {\bs k}_\perp,x)&=&\frac1{2}\Big[\rho^{[\gamma^+]}({\bs b}_\perp, {\bs k}_\perp,x,\hat{\bs e}_i)-\rho^{[\gamma^+]}({\bs b}_\perp, {\bs k}_\perp,x,-\hat{\bs e}_i)\Big]
\ea
The transverse-longitudinally polarized  Wigner distribution
\ba
\rho^i_{TL}({\bs b}_\perp, {\bs k}_\perp,x)&=&\frac1{2}\Big[\rho^{[\gamma^+\gamma^5]}({\bs b}_\perp, {\bs k}_\perp,x,\hat{\bs e}_i)-\rho^{[\gamma^+\gamma^5]}({\bs b}_\perp, {\bs k}_\perp,x,-\hat{\bs e}_i)\Big]
\ea
The transversely polarized  Wigner distribution
\ba
\rho_{TT}({\bs b}_\perp, {\bs k}_\perp,x)&=&\frac1{2}\delta_{ij}\Big[\rho^{[i \sigma^{+j}\gamma^5]}({\bs b}_\perp, {\bs k}_\perp,x,\hat{\bs e}_i)-\rho^{[i \sigma^{+j}\gamma^5]}({\bs b}_\perp, {\bs k}_\perp,x,-\hat{\bs e}_i)\Big]
\label{rhott}
\ea
For $i=j=1$ and $i=j=2$, the result is the same.\\
The pretzelous  Wigner distribution 
\ba\label{pretz}
\rho^{\perp}_{TT}({\bs b}_\perp, {\bs k}_\perp,x)&=& \frac1{2}\epsilon_{ij}\Big[\rho^{[i \sigma^{+j}\gamma^5]}({\bs b}_\perp, {\bs k}_\perp,x,\hat{\bs e}_i)-\rho^{[i \sigma^{+j}\gamma^5]}({\bs b}_\perp, {\bs k}_\perp,x,-\hat{\bs e}_i)\Big]
\ea
where 
$+ \hat{\bs e}_z$ ($- \hat{\bs e}_z$) corresponds to helicity up (down), i.e, $\vert  \frac1{2} \rangle$ ($\vert  -\frac1{2} \rangle$) of the target state. 
$\hat{\bs e}_i$ corresponds to the transversity state and can be expressed in terms of the helicity state. For instance, 
$\vert\pm \hat{\bs e}_x\rangle=\frac1{\sqrt{2}}(\vert \frac1{2} \rangle\pm\vert -\frac1{2} \rangle)$.\\
Here Equation(\ref{pretz}) corresponds to the transversely polarized dressed quark 
and internal quark along the two orthogonal directions. For example, consider the case, 
$i=1$, $j=2$, that refers to the dressed quark  polarized along the $x-$direction and 
the internal quark polarized along the $y-$direction. There are two terms corresponding to this 
case as seen on the rhs of Equation (\ref{pretz}). We obtain an equal contribution from both terms. Thus,  for case $i=1$, $j=2$ the overall contribution to 
$\rho^{\perp}_{TT}({\bs b}_\perp, {\bs k}_\perp,x)$ is zero. Similarly, for the case 
$i=2$, $j=1$, we obtain that the distribution $\rho^{\perp}_{TT}({\bs b}_\perp, {\bs k}_\perp,x)$ 
vanishes. Thus the pretzelous Wigner distribution vanishes in our model. We also would like to point 
out that the pretzelous distribution in Ref. \cite{Liu14} vanishes in the scalar spectator case
but not in the axial-vector spectator case.
In this model, we obtained $\rho_{UL}({\bs b}_\perp, {\bs k}_\perp,x)$ equal to $\rho_{LU}({\bs b}_\perp, {\bs k}_\perp,x)$. 
So, we have ten independent Wigner distributions, out of which $\rho^{j}_{TT}({\bs b}_\perp, {\bs k}_\perp,x)$ with $j=1, 2$ 
vanishes as discussed above.

Finally, we study the eight independent Wigner distributions, and their analytical expressions are as follows:
\ba
\rho_{UU}({\bs b}_\perp,{\bs k}_\perp,x)&=& N\!\!\int\!\!\frac{d^2\Delta_{\perp}}{2(2\pi)^2}\frac{\cos({\bs \Delta}_{\perp}\cdot{\bs b}_{\perp})}{D({\bs q}_\perp)D({\bs q'}_\perp)}
\nn\\
 &\times&\Big[\frac{\Big(4k_\perp^2-\Delta_\perp^2 (1-x)^2\Big)(1+x^2)}{x^2(1-x)^3}+\frac{4 m^2(1-x)}{x^2}\Big]
\label{rhouu} 
\ea
\ba
\rho_{UL}({\bs b}_\perp, {\bs k}_\perp,x)&=&N \int \frac{d^2\Delta_{\perp}}{2(2\pi)^2}\, \frac{\sin({\bs \Delta}_{\perp}\cdot {\bs b}_{\perp})}{D({\bs q}_\perp)D({\bs q'}_\perp)}
 \Big[\frac{4\Big(k_y\Delta_x-k_x\Delta_y\Big)(1+x)}{x^2(1-x)}\Big]\label{rhoul}
\ea
\ba
\rho^x_{UT}({\bs b}_\perp, {\bs k}_\perp,x)&=& N \int \frac{d^2\Delta_{\perp}}{2(2\pi)^2}\, \frac{\sin({\bs \Delta}_{\perp}\cdot{\bs b}_{\perp})}{D({\bs q}_\perp)D({\bs q'}_\perp)}
 \Big[\frac{4m \Delta_x}{x^2}\Big]\label{rhout}
\ea
\ba\rho_{LL}({\bs b}_\perp, {\bs k}_\perp,x)&=&N \!\!\int\!\! \frac{d^2\Delta_{\perp}}{2(2\pi)^2}\, \frac{\cos({\bs \Delta}_{\perp}\cdot{\bs b}_{\perp})}{D({\bs q}_\perp)D({\bs q'}_\perp)}\nn\\
&\times& \Big[\frac{\Big(4k_\perp^2-\Delta_\perp^2 (1-x)^2\Big)(1+x^2)}{x^2(1-x)^3}-\frac{4 m^2(1-x)}{x^2}\Big]\label{rholl}
\ea
\ba
\rho^x_{LT}({\bs b}_\perp, {\bs k}_\perp,x)&=&N \int \frac{d^2\Delta_{\perp}}{2(2\pi)^2}\, \frac{\cos({\bs \Delta}_{\perp}\cdot{\bs b}_{\perp})}{D({\bs q}_\perp)D({\bs q'}_\perp)}
  \Big[\frac{8m k_x}{x^2(1-x)}\Big]\label{rholt}
  \ea
\ba
\rho^x_{TU}({\bs b}_\perp, {\bs k}_\perp,x)&=&N \int \frac{d^2\Delta_{\perp}}{2(2\pi)^2}\, \frac{\sin({\bs \Delta}_{\perp}\cdot{\bs b}_{\perp})}{D({\bs q}_\perp)D({\bs q'}_\perp)}
 \Big[\frac{4m \Delta_x}{x}\Big]\label{rhotu}
\ea
\ba
\rho^x_{TL}({\bs b}_\perp, {\bs k}_\perp,x)&=& N \int \frac{d^2\Delta_{\perp}}{2(2\pi)^2}\, \frac{\cos({\bs \Delta}_{\perp}\cdot{\bs b}_{\perp})}{D({\bs q}_\perp)D({\bs q'}_\perp)}
 \Big[\frac{-8m k_x}{x(1-x)}\Big]\label{rhotl}
\ea
\ba\rho_{TT}({\bs b}_\perp, {\bs k}_\perp,x)&=&N \int \frac{d^2\Delta_{\perp}}{2(2\pi)^2}\, \frac{\cos({\bs \Delta}_{\perp}\cdot{\bs b}_{\perp})}{D({\bs q}_\perp)D({\bs q'}_\perp)}
  \Big[\frac{2\Big(4k_\perp^2-\Delta_\perp^2 (1-x)^2\Big)}{x(1-x)^3}\Big]\label{rhott}
\ea
where,
$$N=\frac{g^2 C_F}{(2 \pi)^3},~~C_F \text{ is the color factor}$$
\ba 
D({\bs q}_\perp)=\Big{[ }   m^2 - \frac{m^2 + ({\bs k_\perp}+\frac{{\bs \Delta_\perp} (1-x)}{2})^2 }{x} - \frac{({\bs k_\perp}+\frac{{\bs \Delta_\perp} (1-x)}{2})^2}{1-x} \Big]\nn\\
D({\bs q}'_\perp)=\Big{[ }   m^2 - \frac{m^2 + ({\bs k_\perp}-\frac{{\bs \Delta_\perp} (1-x)}{2})^2 }{x} - \frac{({\bs k_\perp}-\frac{{\bs \Delta_\perp} (1-x)}{2})^2}{1-x} \Big]
\ea
\section{Numerical strategy}\label{numerical}

The eight independent Wigner distributions obtained in the previous section are function of 
five continuous variables, two transverse position ${\bs b}_\perp$, two transverse momentum 
${\bs k}_\perp$ and one longitudinal momentum fraction $x$. We are interested in studying the transverse 
phase space, so we integrate over the $x$ dependence from all the distributions and purely 
study them in the transverse space. This integration over the longitudinal 
momentum fraction ought to go from $0$ to $1$ in this model. However, in order to correctly calculate the 
 contribution at $x=1$ we need to incorporate the contribution from the single-particle sector of the Fock space expansion.  
 
This will contribute to $\rho_{UU}$, $\rho_{LL}$, and $\rho_{TT}$. At $O(g^2)$, this part 
gets a contribution from the normalization of the state \cite{Hari99}. The single-particle contribution 
to the Wigner function is of the form $\rho_0({\bs b}_\perp, {\bs k}_\perp,x)=N\, \delta(1-x)\, \delta^2({\bs b}_\perp)\, \delta^2({\bs k}_\perp)$.
This is because it represents a single quark carrying all the momentum at ${\bs b}_\perp=0$ and 
the average transverse momentum is also zero. The delta function peak at ${\bs b}_\perp=0$ gets 
smeared by the contribution from the two-particle sector.

As discussed above, the single-particle contribution to the Wigner distribution 
 corresponds to a single quark carrying all the momentum at ${\bs b}_\perp={\bs k}_\perp=0$. 
In our study, the Wigner distribution does not get the contribution from the  single-particle sector
as we fix a nonzero value for ${\bs b}_\perp$(${\bs k}_\perp$) in ${\bs k}_\perp$(${\bs b}_\perp$) space
which makes $\rho_0({\bs b}_\perp, {\bs k}_\perp,x)=0$.

Now, by fixing the transverse momentum and integrating the longitudinal momentum fraction $x$
from [0, 1], we can study the distributions as functions of $b_x$ and $b_y$ in ${\bs b}_\perp$ space. 
The numerical integration over $x$ from [0, 1] is performed for a very high precision up to $O(10^{-24})$
for the upper limit of $x$ integration. We would like to mention here that only in the ${\bs b}_\perp$ space we observe qualitative difference 
in the results for integration over $x$ from [0, 0.9] versus $x$ [0, 1]  with $x \approx 1$. 
Thus for ${\bs b}_\perp$ space we integrate over $x$ [0, 1] with $x \approx 1$.

Similarly, by fixing the transverse position, we can study the distributions as functions of $k_x$ and $k_y$ in ${\bs k}_\perp$ space. 
In this case also we can integrate $x$ from [0, 1] as mentioned before, but here we observe a very sharp negative peak at the center ($k_x = k_y = 0$) for $\rho_{UU}$, 
$\rho_{LL}$, $\rho_{TT}$, $\rho^x_{UT}$ and $\rho^x_{TU}$ . 
The magnitude of this peak is so large that remaining part of the distribution is not perceived. In order to study
the distributions in ${\bs k}_\perp$ space, the nature of the integrand mandates that
we take the cutoff on 
upper limit of $x$, which enables us to observe feasible distribution in ${\bs k}_\perp$ space. So we choose the upper limit of the 
$x$ integration as $0.9$ for all distributions to study the ${\bs k}_\perp$ space. We would like to highlight the fact
that the qualitative behavior after putting the cutoff $x$ [0, 0.9] is exactly the same as integrating $x$ [0, 1] with 
the upper cutoff $x \approx 1$ . 
We curtail the peak at the origin by putting the cutoff so that we 
can study 
the qualitative behavior of the distributions.

One can also study the mixed space distribution by further integrating out $b_y$ and $k_x$ and plot the distributions as 
a function of the remaining variables, i.e., $b_x$ and $k_y$.
While studying distributions in mixed space we have a a similar situation
 as we observed in ${\bs k}_\perp$ space.
 Thus, in mixed space also we use the same cutoff of 0.9 and obtain the same qualitative behavior as for the one
with the upper cutoff $x \approx 1$. Thus, both in ${\bs k}_\perp$ space and mixed space we obtain the same qualitative
behavior for integration over $x$ from [0, 0.9] and $x$ [0, 1]  with $x \approx 1$.

 The mixed space plots are not subject to Heisenberg's uncertainty condition 
and can be interpreted as probability densities.
\begin{figure}[h]
 \centering 
(a)  \includegraphics[width=7.5cm,height=6.0cm]{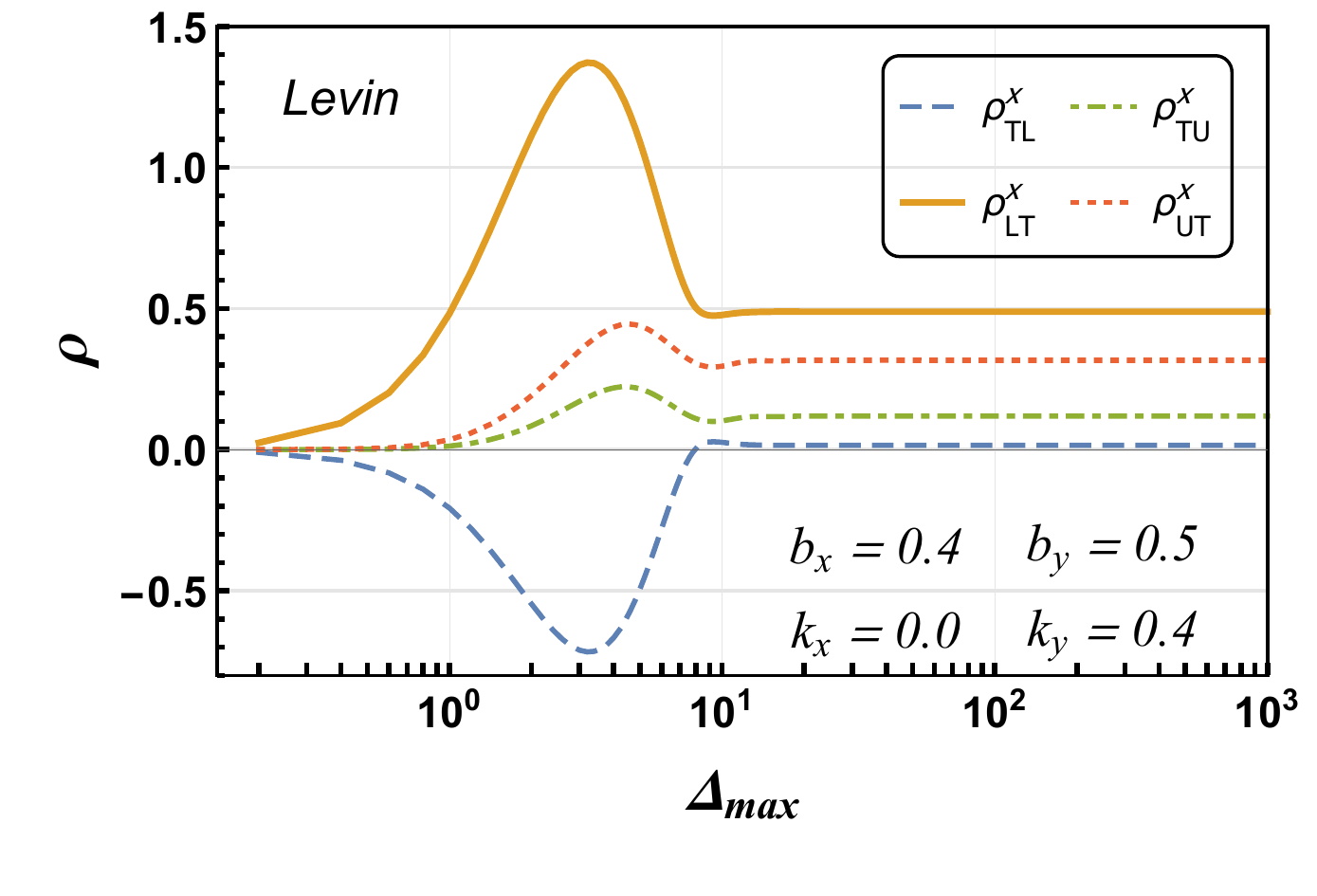}   (c)\includegraphics[width=7.5cm,height=6.0cm]{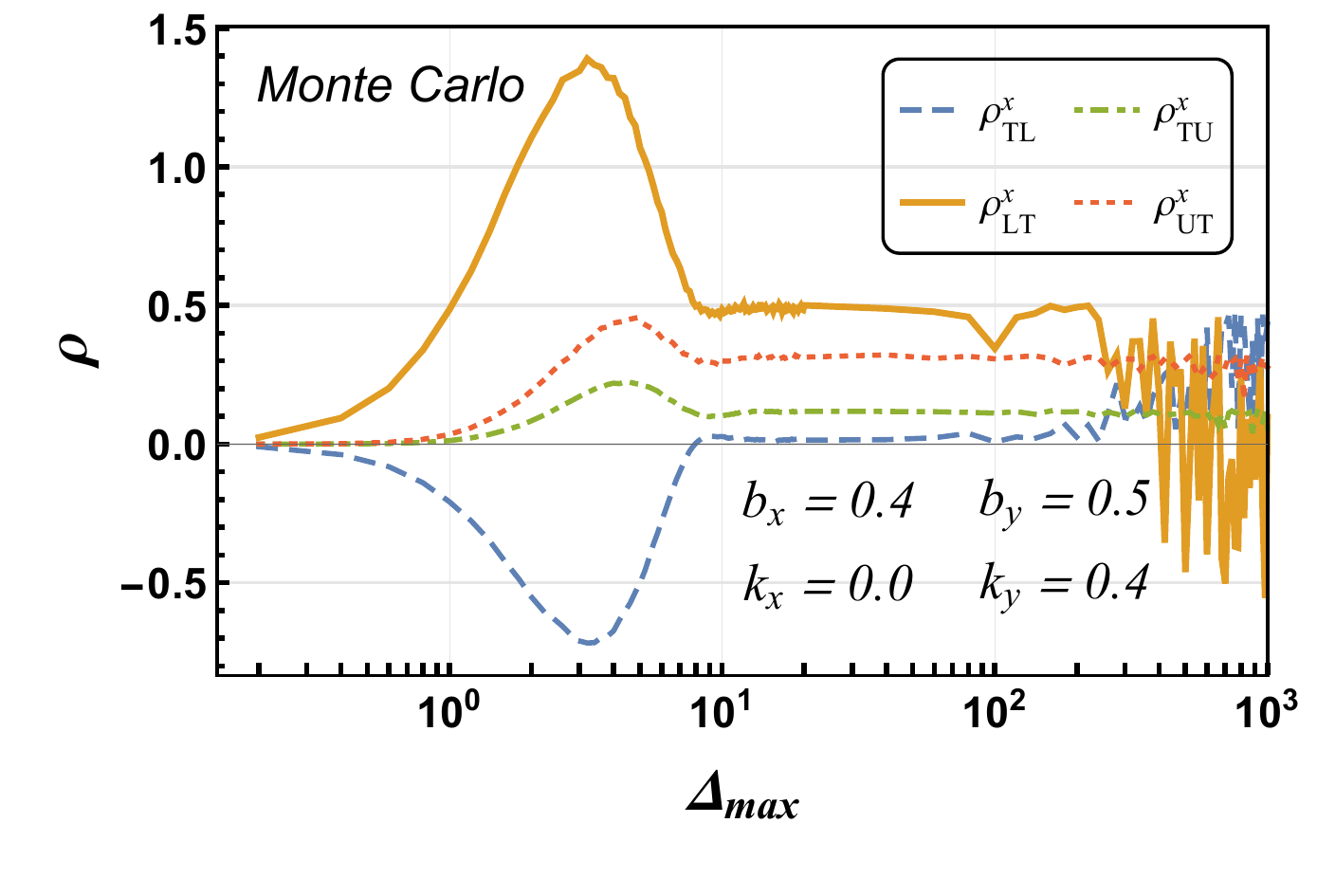}\\[2ex]
(b)\includegraphics[width=7.5cm,height=6.0cm]{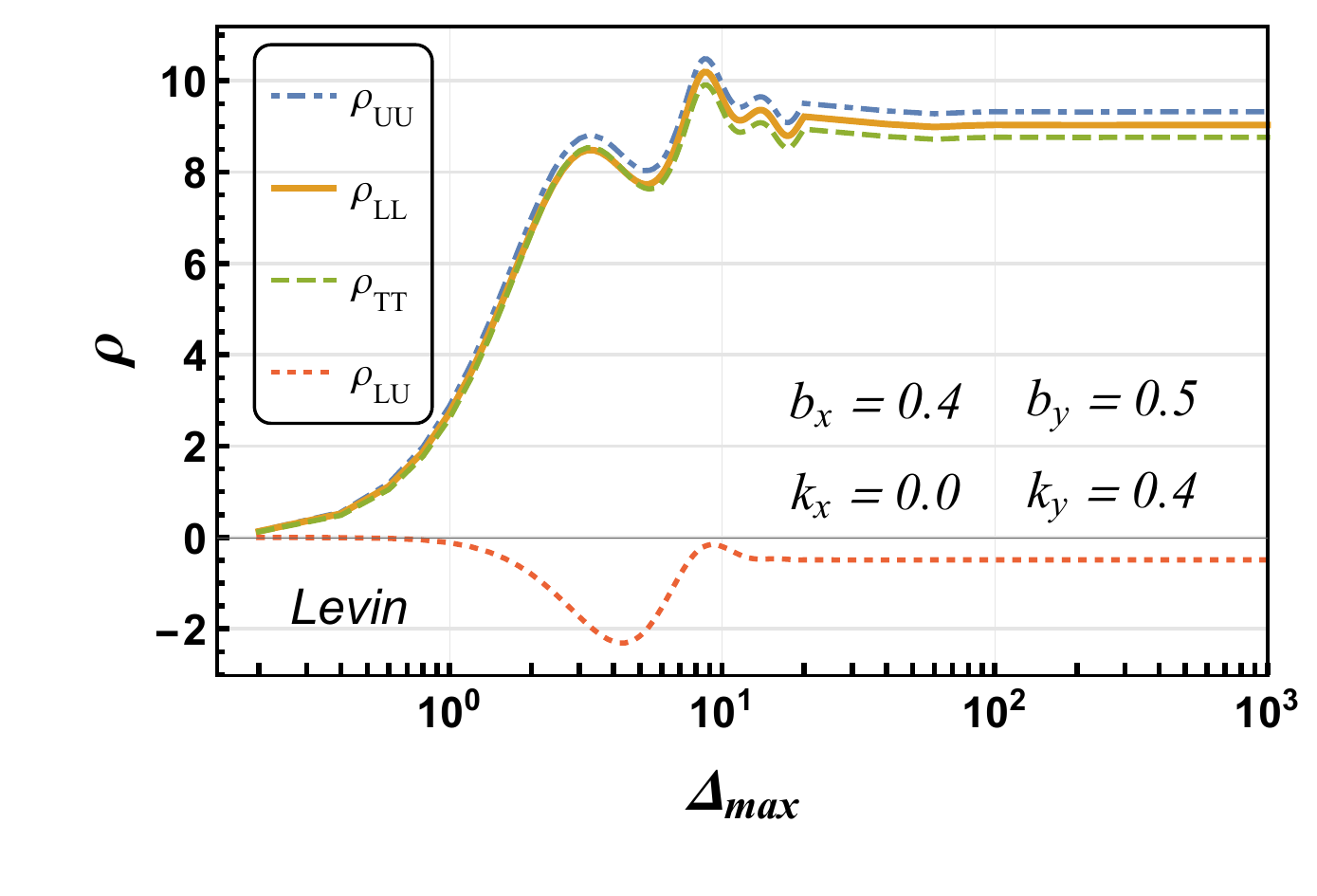}   (d)\includegraphics[width=7.5cm,height=6.0cm]{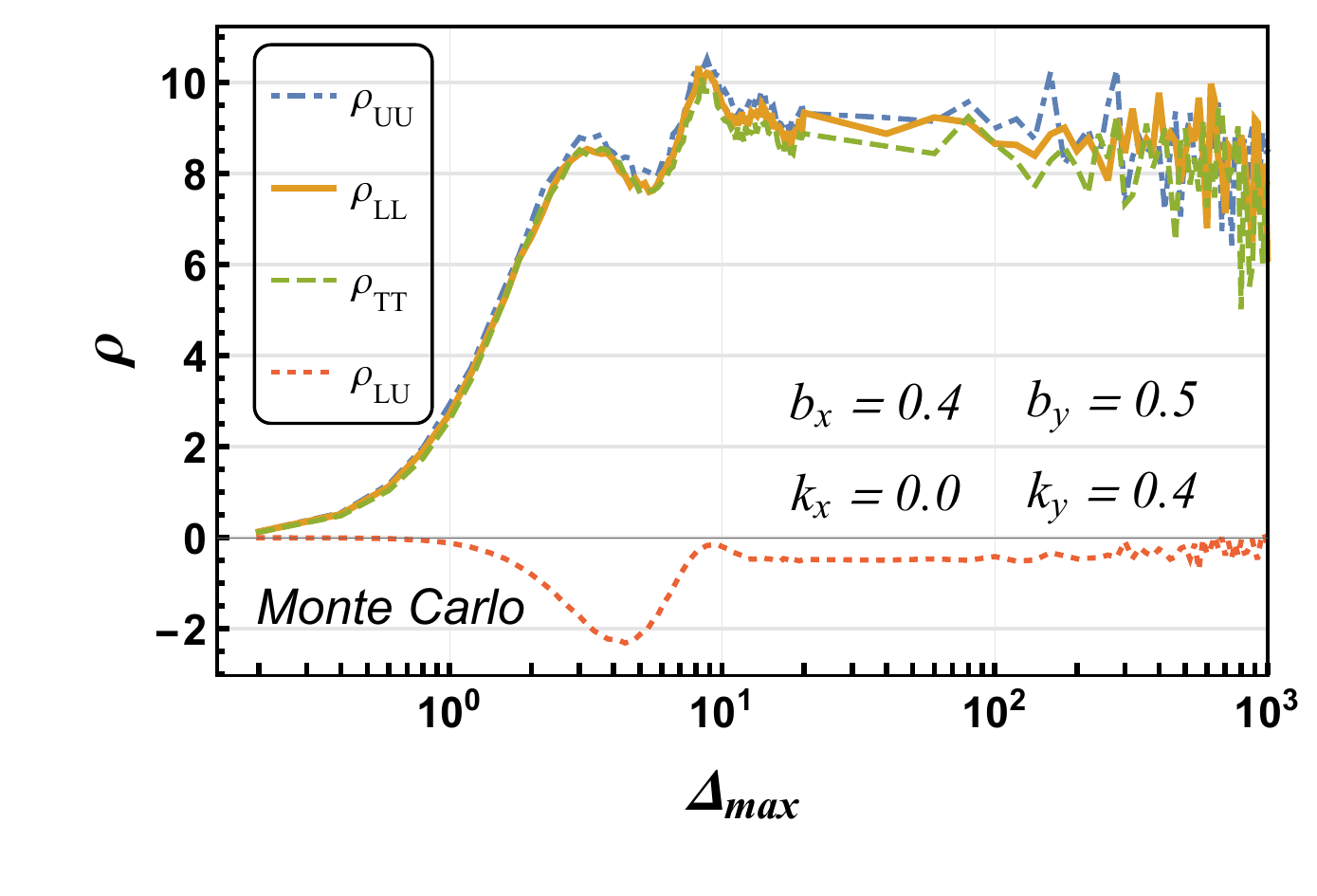}
\caption{Plot of eight Wigner distributions vs $\Delta_{max}(\mathrm{GeV})$ at a fixed value of $b_x~(\mathrm{GeV}^{-1})$, $b_y~(\mathrm{GeV}^{-1})$, 
$k_x~(\mathrm{GeV})$, and $k_y~(\mathrm{GeV})$ using the Levin and Monte Carlo integration methods. 
Plots (a) and (b) are generated by using the Levin method. Plots (c) and (d) are generated by using the Monte Carlo method.}
  \label{fig0}
\end{figure}
The Fourier transform in the definition of a Wigner distribution involves doing an integration over ${\bs \Delta}_{\perp}$ which ideally 
should go from $-\infty$ to $\infty$, but since we are performing a numerical calculation we have to choose a suitable cutoff. 
Since our integrand involves highly oscillatory function we use the Levin method for doing the numerical integration, which is tailor-made for functions that have oscillatory behavior. The Levin method gives us converging results as compared to 
conventional numerical methods like Monte Carlo (MC). In our previous work \cite{Asmita14} on Wigner distributions we had relied on MC integration and thus the results were cutoff dependent. 
%
However, for lower values of $\Delta_{max}$  both the MC and Levin methods are in good 
agreement with each other (see Figure.~\ref{fig0}).
%

In Figure.~\ref{fig0} we show the behavior of all the Wigner distributions with $\Delta_{max}$ using two methods for numerical integration, i.e., the  
Levin and MC methods. In Figures.~\ref{fig0} (a) and \ref{fig0}(c) we show the 
distributions $\rho^x_{TL},\,\rho^x_{TU},\,\rho^x_{LT}$, and $\rho^x_{UT}$ for the Levin and MC methods, respectively.
Similarly in Figures.~\ref{fig0} (b) and \ref{fig0}(d) we show the distributions for $\rho_{UU},\,\rho_{LL},\,\rho_{TT}$, and $\rho_{LU}$. The 2D plots are for a fixed value of $b_x=0.4~\mathrm{GeV}^{-1}$ , $b_y=0.5~\mathrm{GeV}^{-1}$, $k_x=0.0~\mathrm{GeV}$, and  $k_y=0.4~\mathrm{GeV}$.
We study the $\Delta_{max}$ dependence up to $\Delta_{max}=1000~\mathrm{GeV}$ and the results clearly show that the
Levin method is ideal since it shows convergence, whereas MC fails to converge. As $\Delta_{max}$ increases,
the results from MC begin to diverge more and more. On the contrary, the Levin method starts to give constant
results from around $\Delta_{max} = 20~\mathrm{GeV}$ and it stays constant thereafter.
Based on these results we set $\Delta_{max} = 20~\mathrm{GeV}$ for all the 3D plots.
In all the plots we have taken $m=0.33~\mathrm{GeV}$, and divided by a normalization constant.

\begin{figure}[h] 
 \centering 
(a)  \includegraphics[width=7.3cm,height=4.7cm]{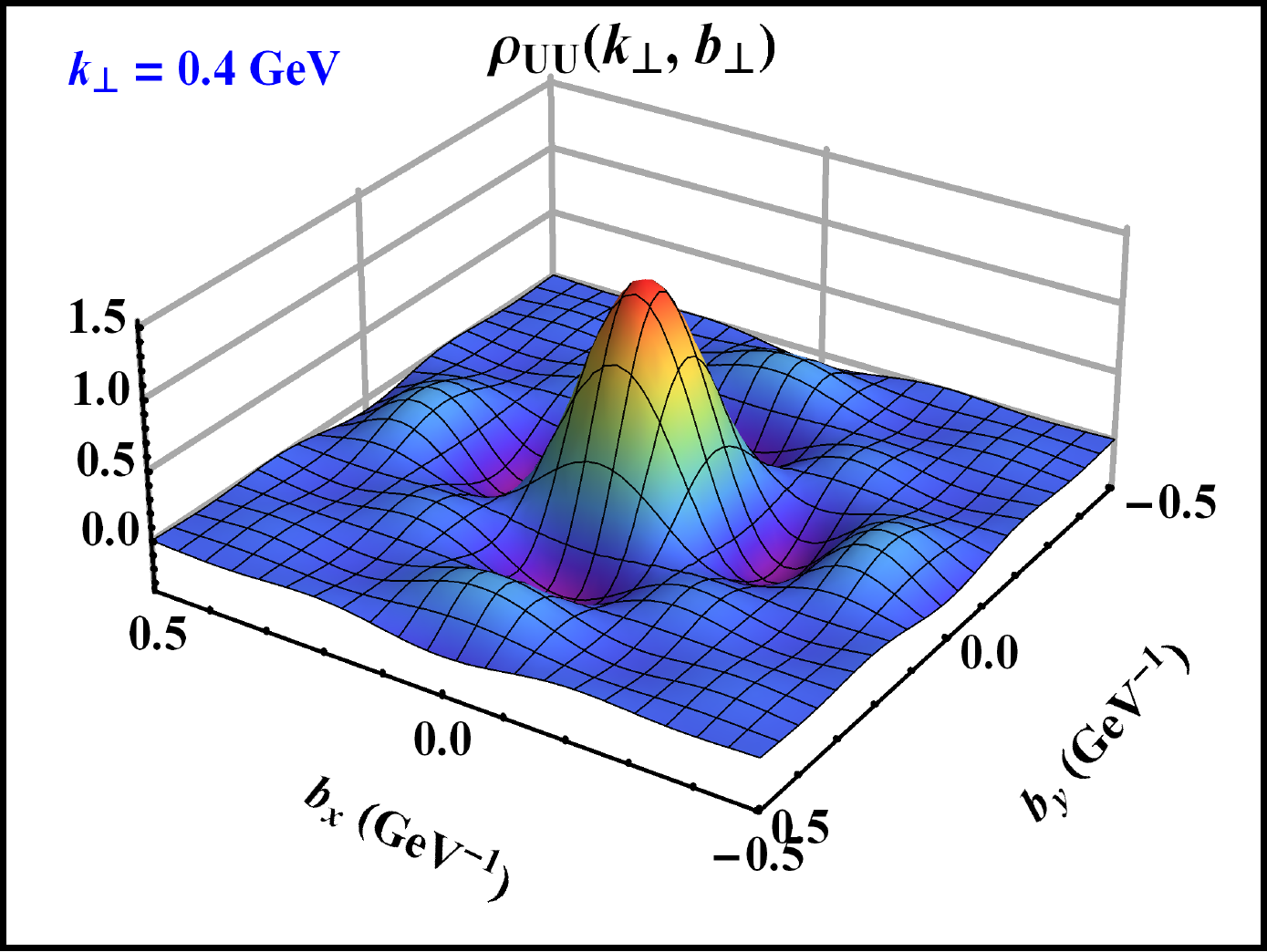}\hskip 0.3cm(d)\includegraphics[width=7.3cm,height=4.7cm]{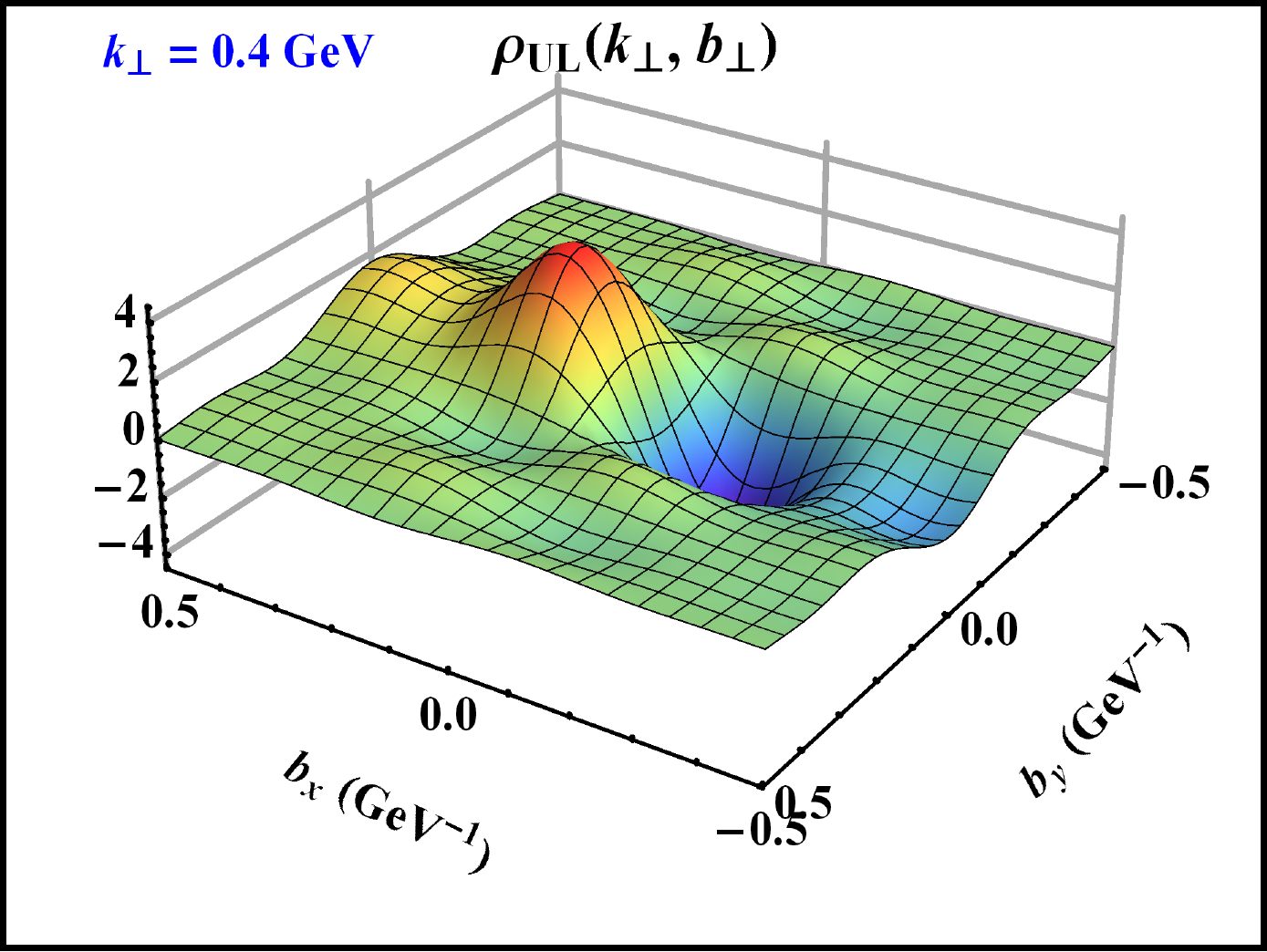}\\[5ex]
(b)\includegraphics[width=7.3cm,height=4.7cm]{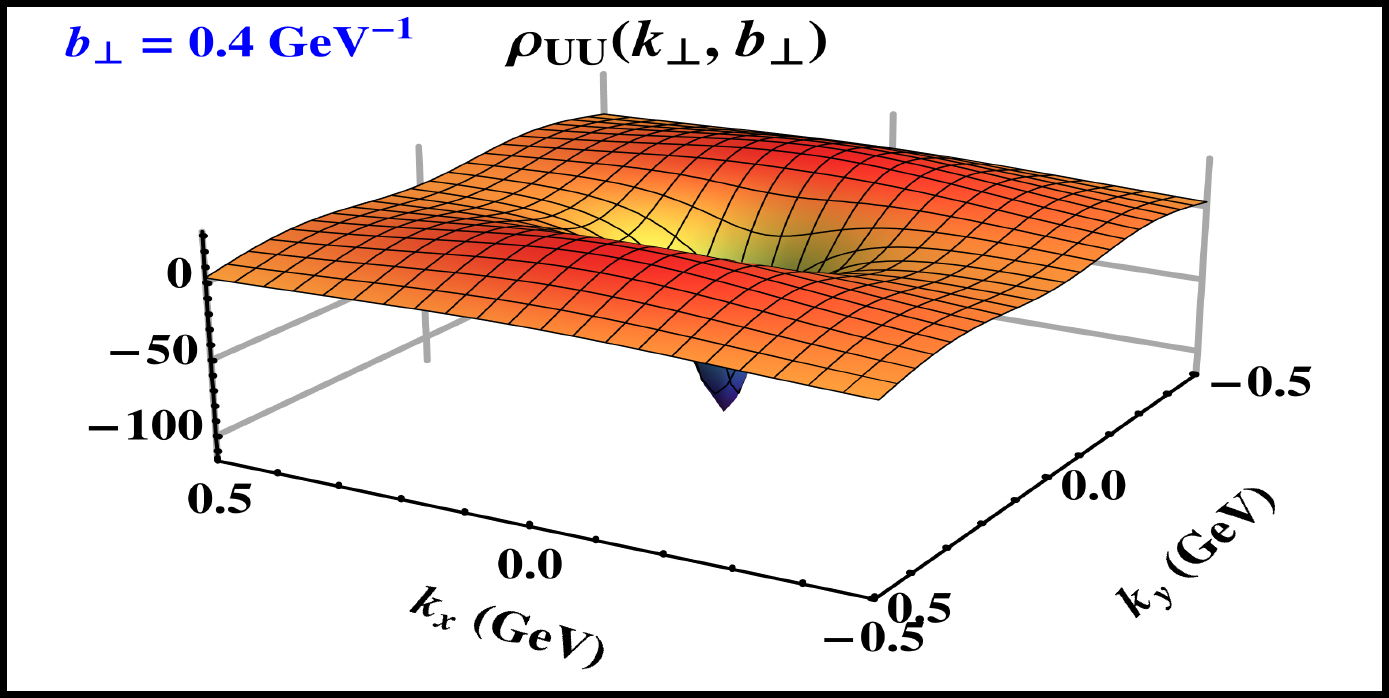}\hskip 0.3cm(e)\includegraphics[width=7.3cm,height=4.7cm]{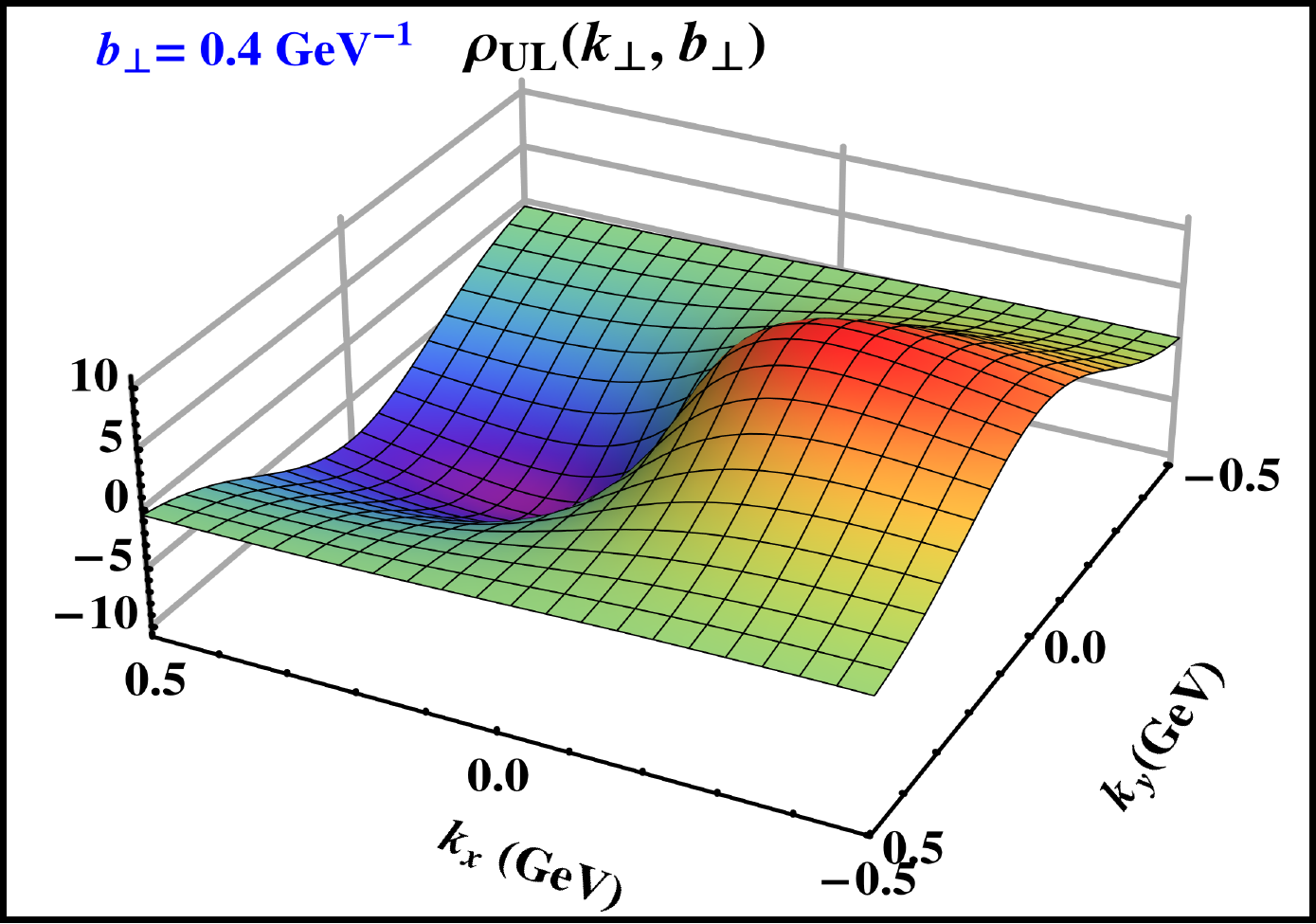}\\[5ex]
(c) \includegraphics[width=7.3cm,height=4.7cm]{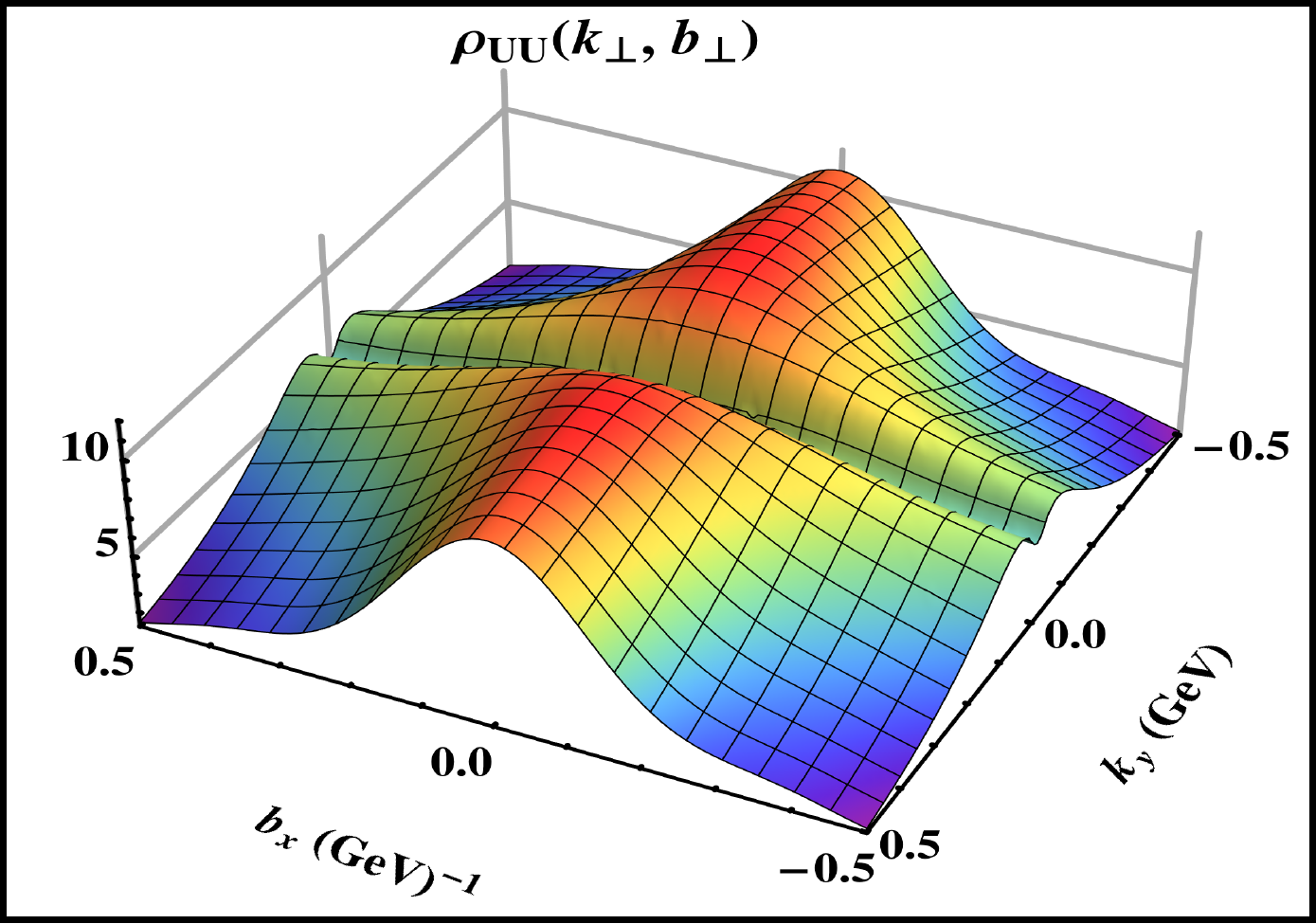}\hskip 0.3cm(f)\includegraphics[width=7.3cm,height=4.7cm]{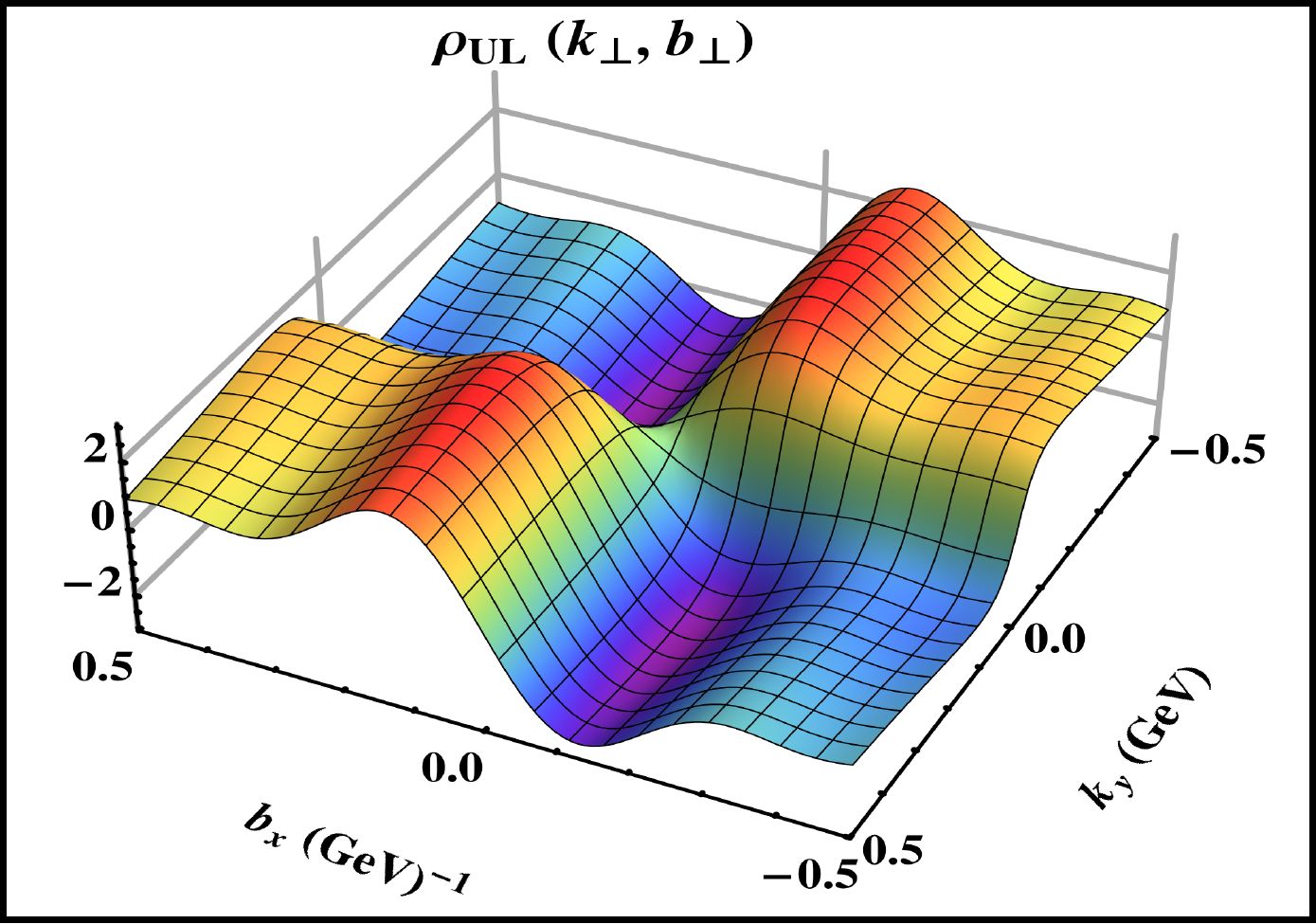}  
  \caption{3D plot of  Wigner distributions $\rho_{UU}({\bs k}_\perp, {\bs b}_\perp)$ and 
  $\rho_{UL}({\bs k}_\perp, {\bs b}_\perp)$ at $\Delta_{max} = 20~\mathrm{GeV}$. The first row displays the two distributions
  in ${\bs b}_\perp$ space with ${\bs k}_\perp= 0.4~\mathrm{GeV} \,\hat{{\bs e}}_y$ and $\rho_{UU}\,(\rho_{UL})$ is scaled by a factor
  $10^{-5}\,(10^{-1})$. The second row shows the two distributions
  in ${\bs k}_\perp$ space with ${\bs b}_\perp= 0.4~\mathrm{GeV}^{-1}\, \hat{{\bs e}}_y$.
  The last row represents the two distributions in mixed space.}  \label{rhou}
\end{figure}
\begin{figure}[h]
 \centering
(a)  \includegraphics[width=7.3cm,height=4.7cm]{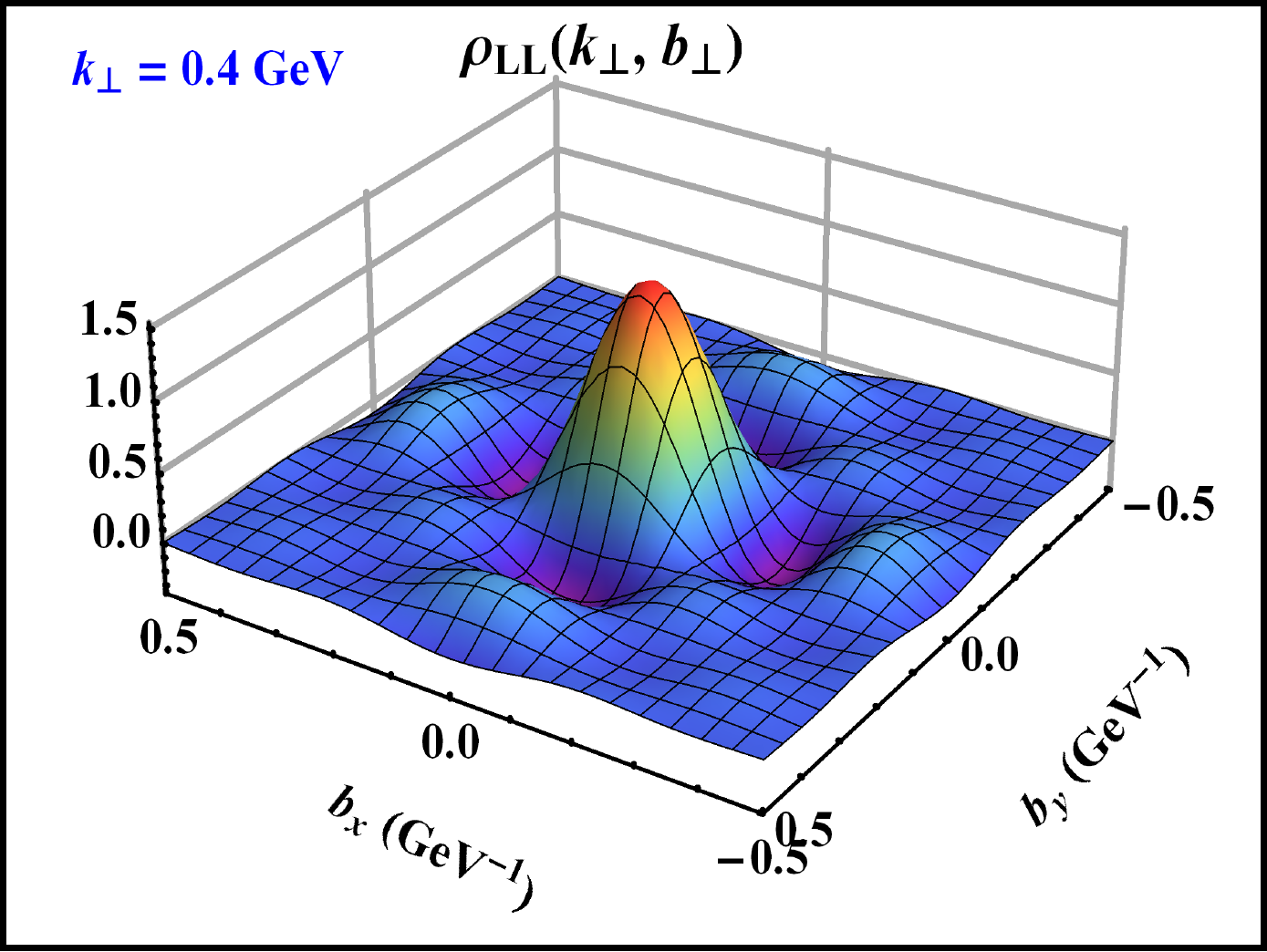}\hskip 0.3cm(d)\includegraphics[width=7.3cm,height=4.7cm]{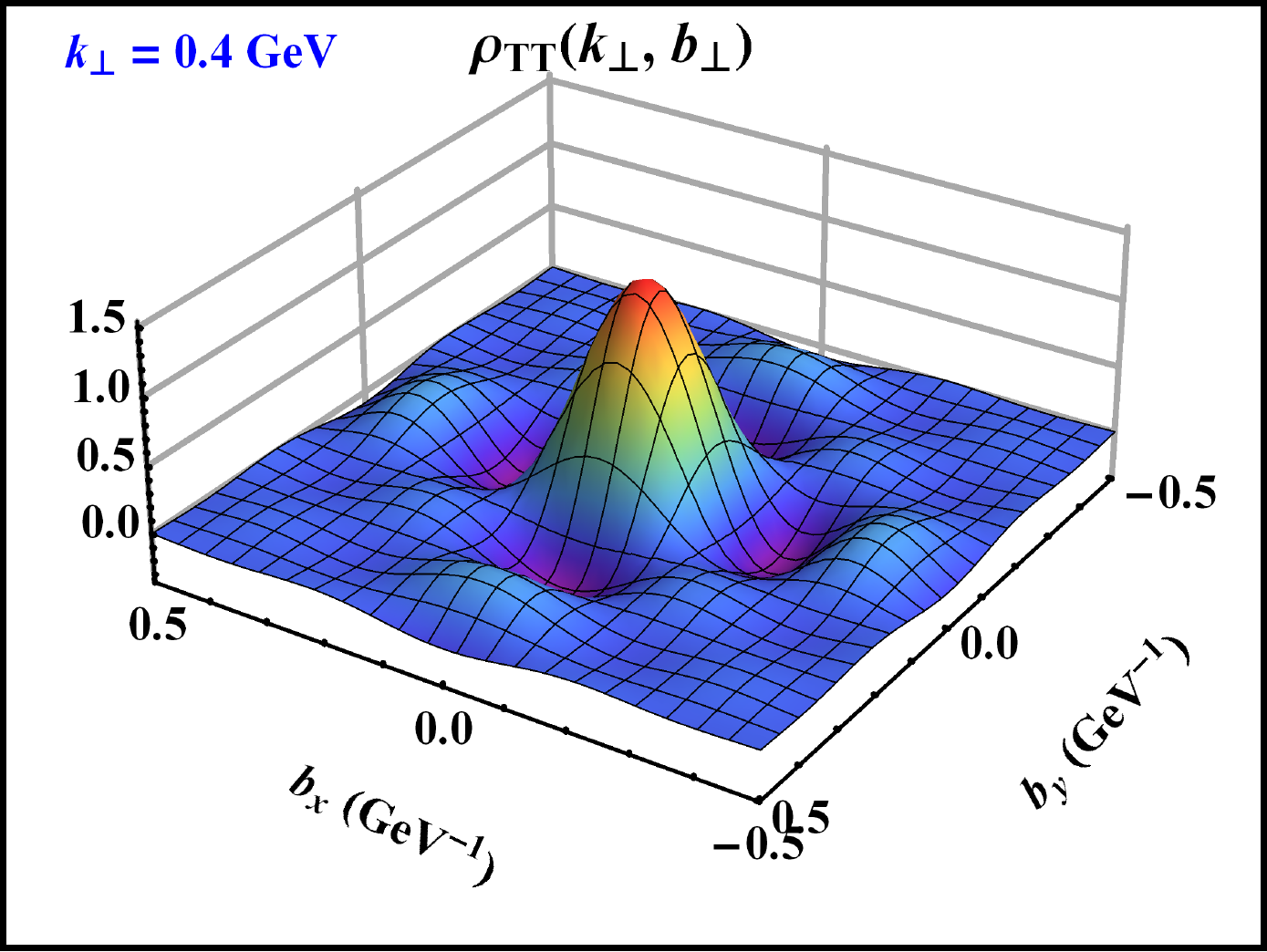}\\[5ex]
(b)\includegraphics[width=7.3cm,height=4.7cm]{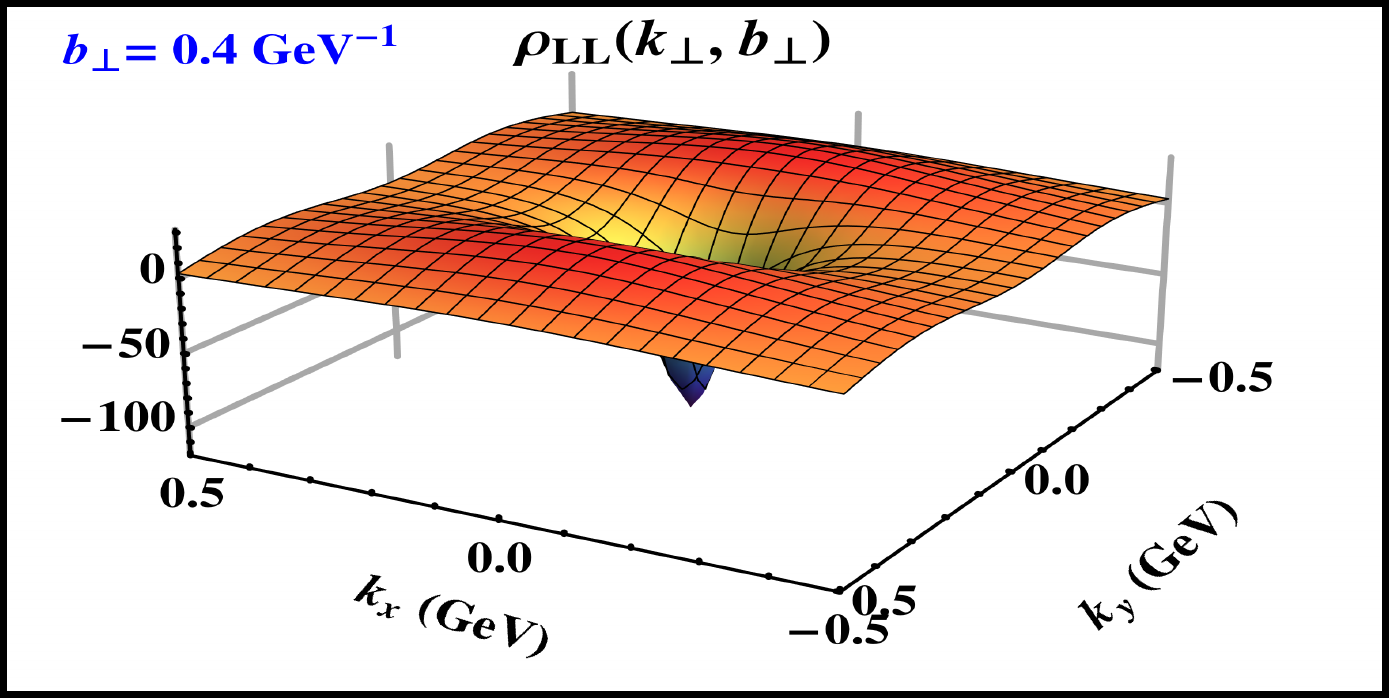}\hskip 0.3cm(e)\includegraphics[width=7.3cm,height=4.7cm]{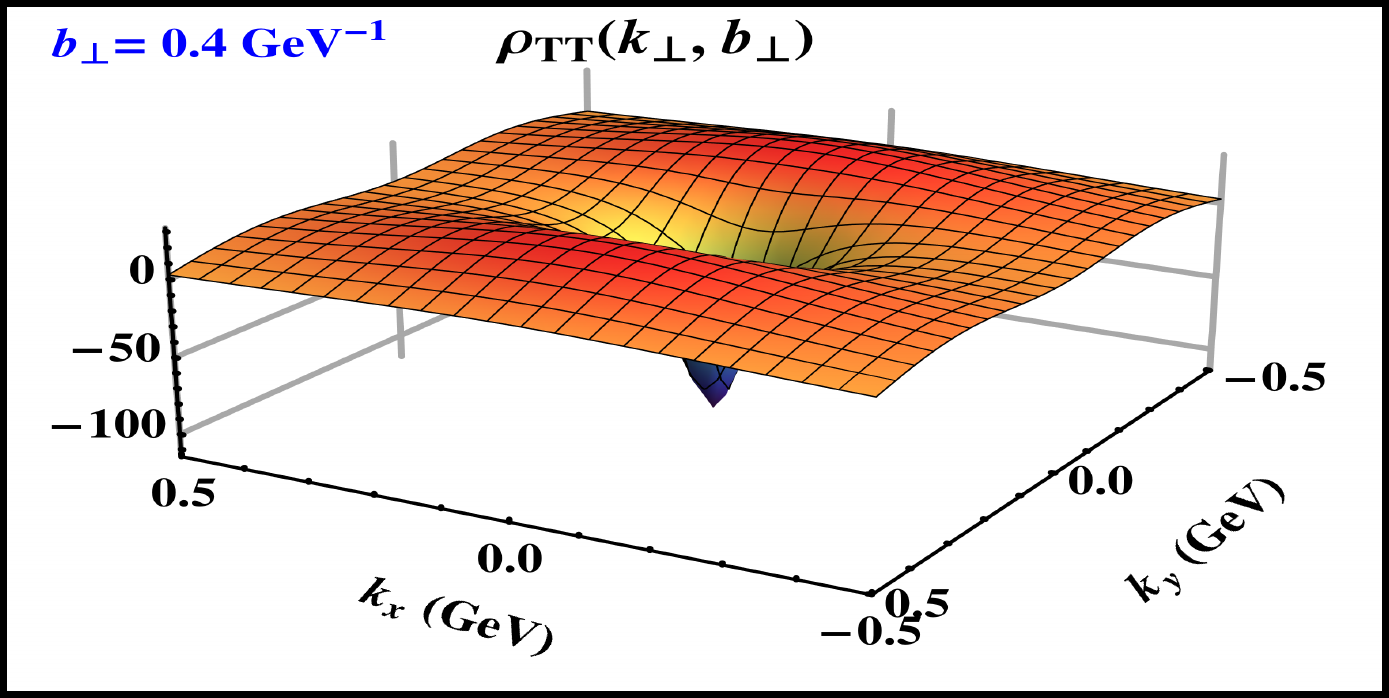}\\[5ex]
(c) \includegraphics[width=7.3cm,height=4.7cm]{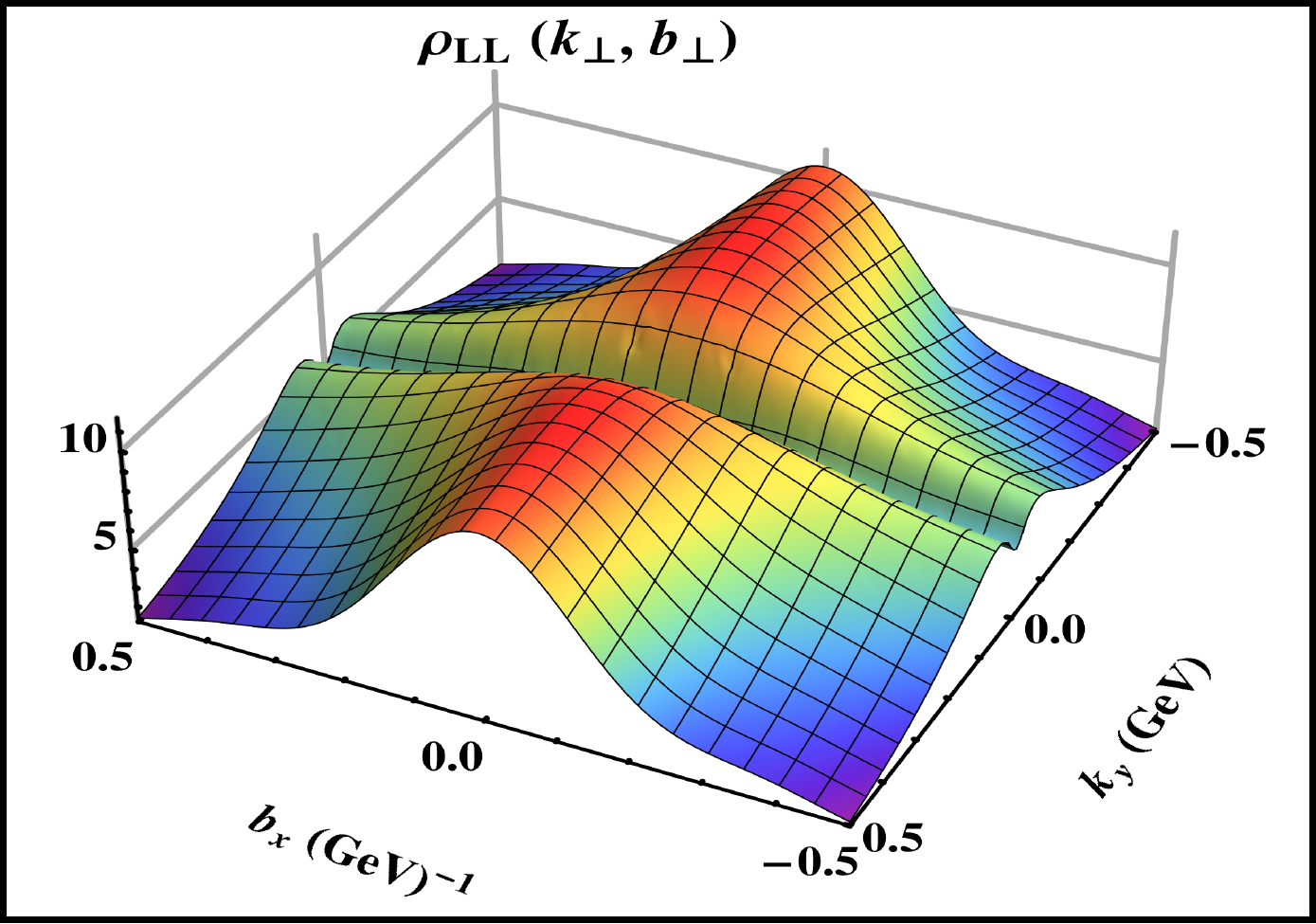}\hskip 0.3cm(f)\includegraphics[width=7.3cm,height=4.7cm]{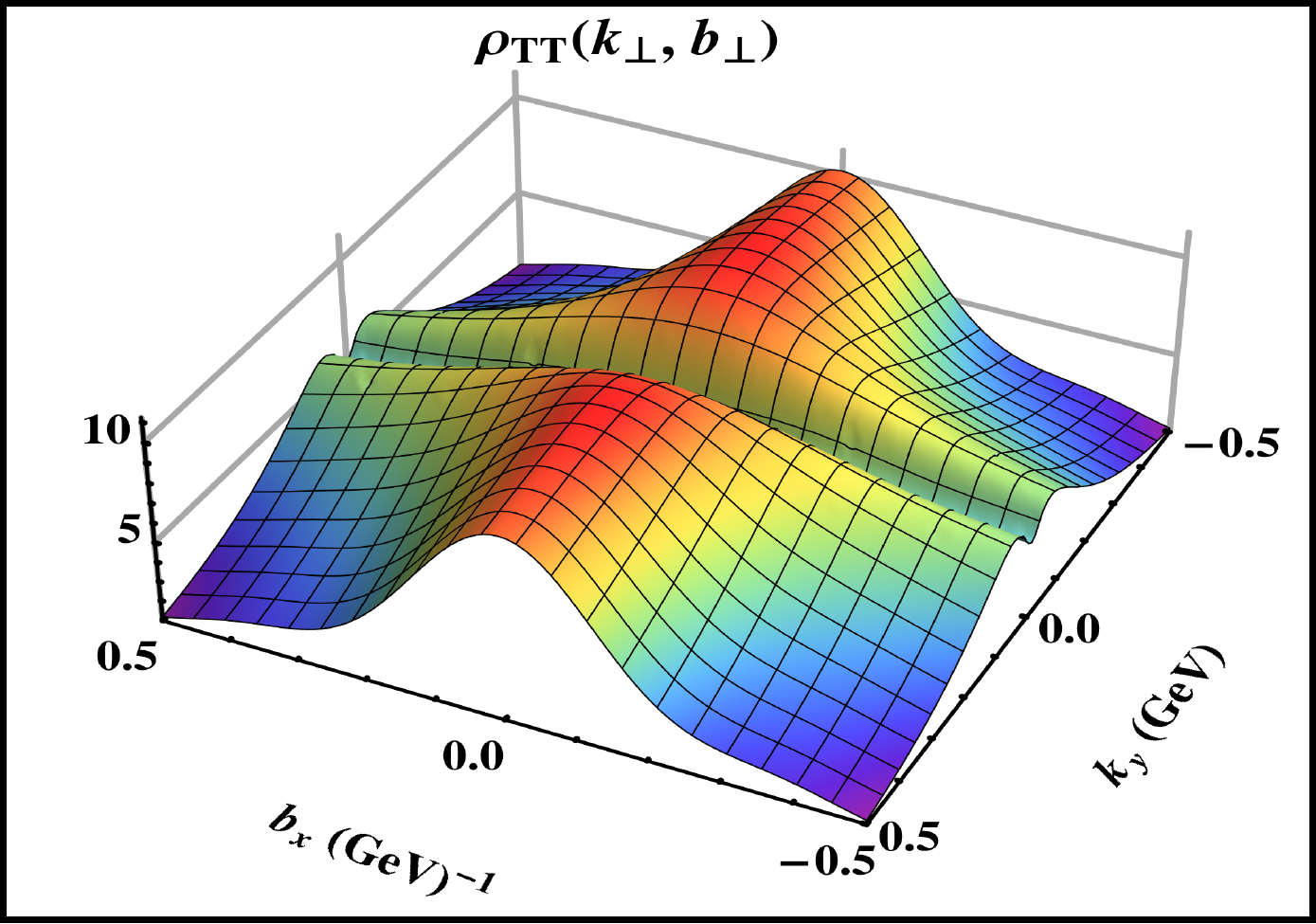}  
  \caption{3D plot of Wigner distributions $\rho_{LL}({\bs k}_\perp, {\bs b}_\perp)$ and 
  $\rho_{TT}({\bs k}_\perp, {\bs b}_\perp)$ at $\Delta_{max} = 20~\mathrm{GeV}$. The first row displays the two distributions
  in ${\bs b}_\perp$ space with ${\bs k}_\perp= 0.4~\mathrm{GeV} \,\hat{{\bs e}}_y$ and $\rho_{LL}$ and $\rho_{TT}$ are 
  scaled by a factor $10^{-5}$. The second row shows the two distributions
  in ${\bs k}_\perp$ space with ${\bs b}_\perp= 0.4~\mathrm{GeV}^{-1}\, \hat{{\bs e}}_y$. The last row represents the two distributions in mixed space.}
 \label{rhol}
\end{figure}
\begin{figure}[h]
 \centering
(a)  \includegraphics[width=7.3cm,height=4.7cm]{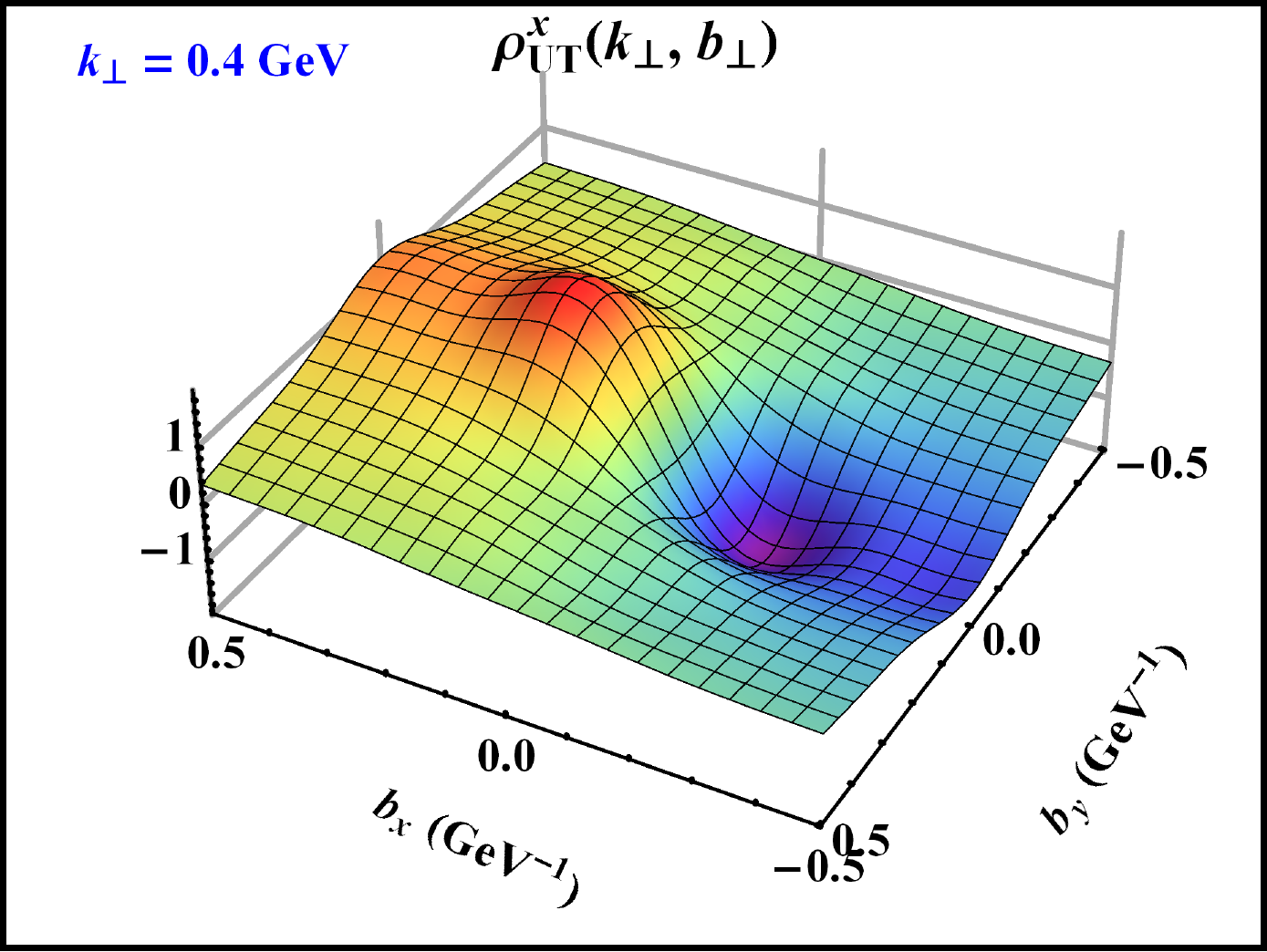}\hskip 0.3cm(d)\includegraphics[width=7.3cm,height=4.7cm]{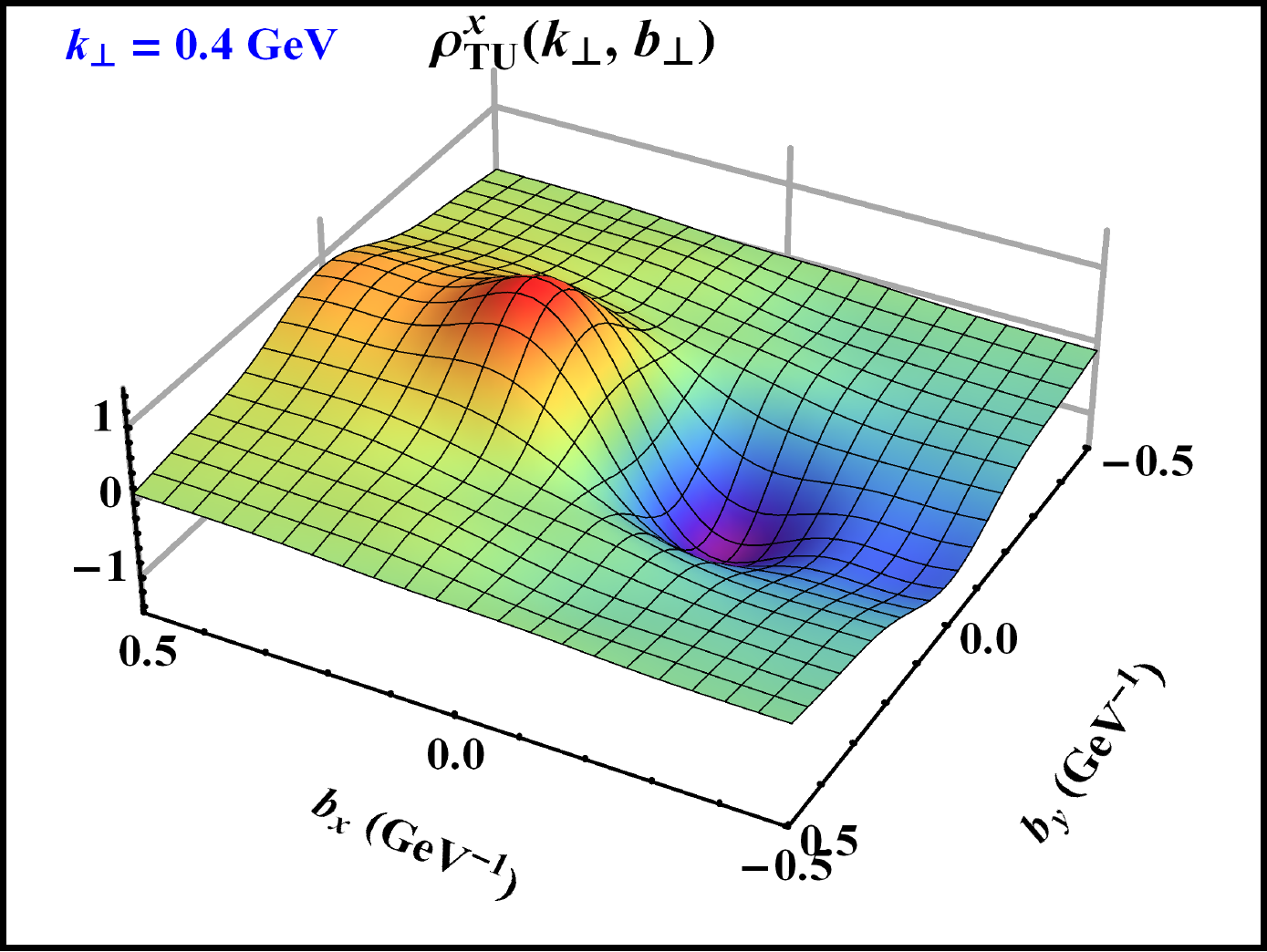}\\[5ex]
(b)\includegraphics[width=7.3cm,height=4.7cm]{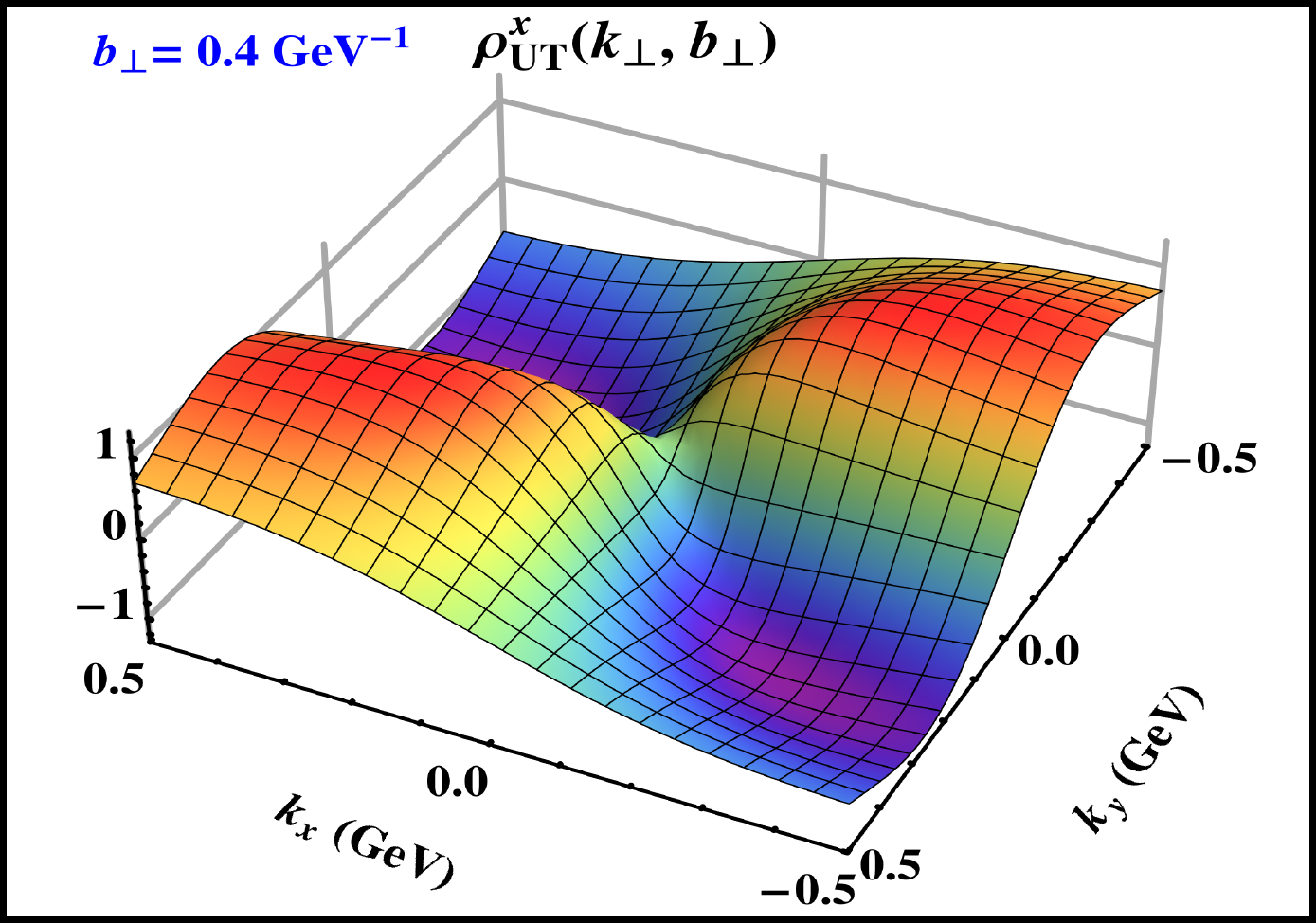}\hskip 0.3cm(e)\includegraphics[width=7.3cm,height=4.7cm]{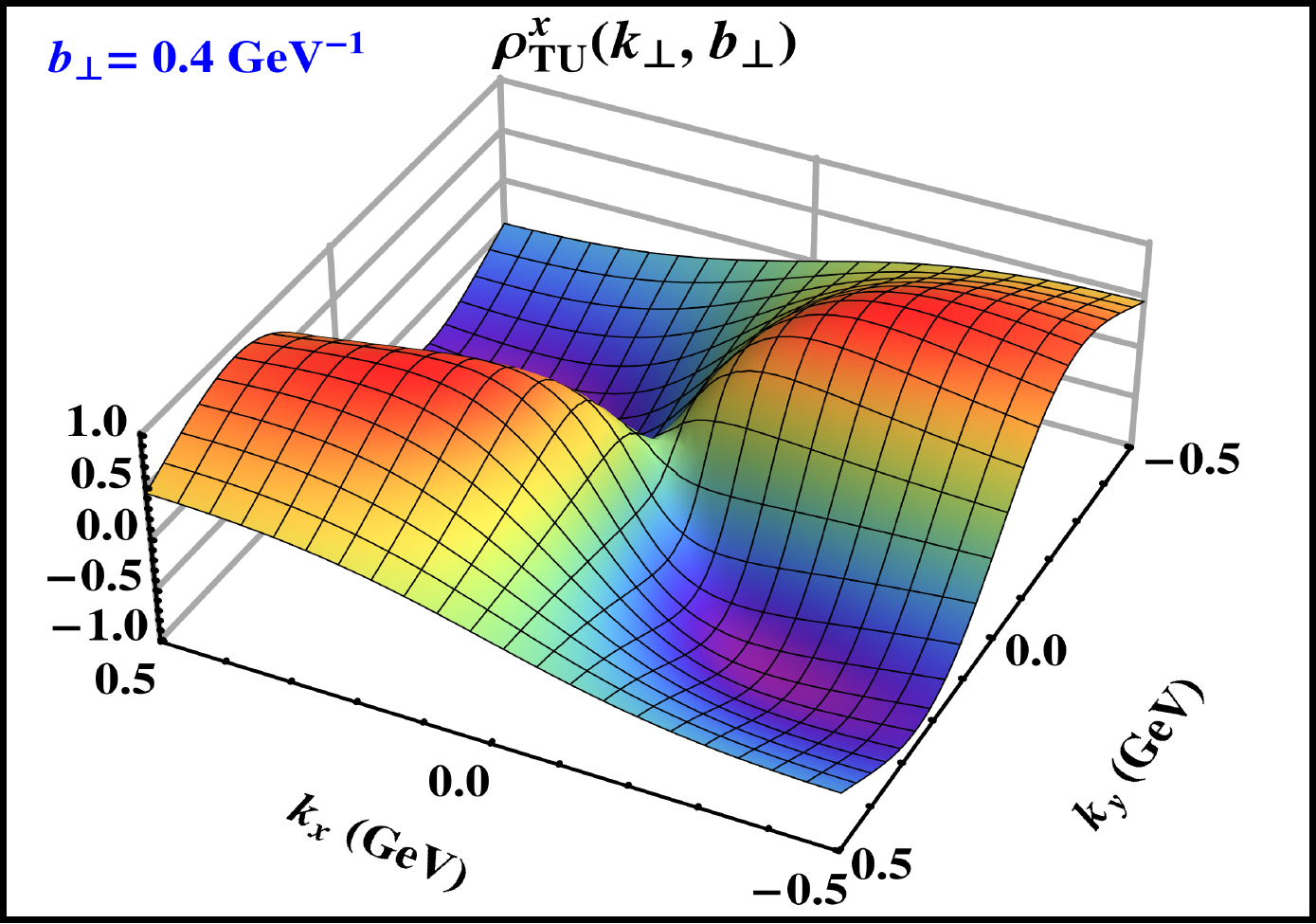}\\[5ex]
(c) \includegraphics[width=7.3cm,height=4.7cm]{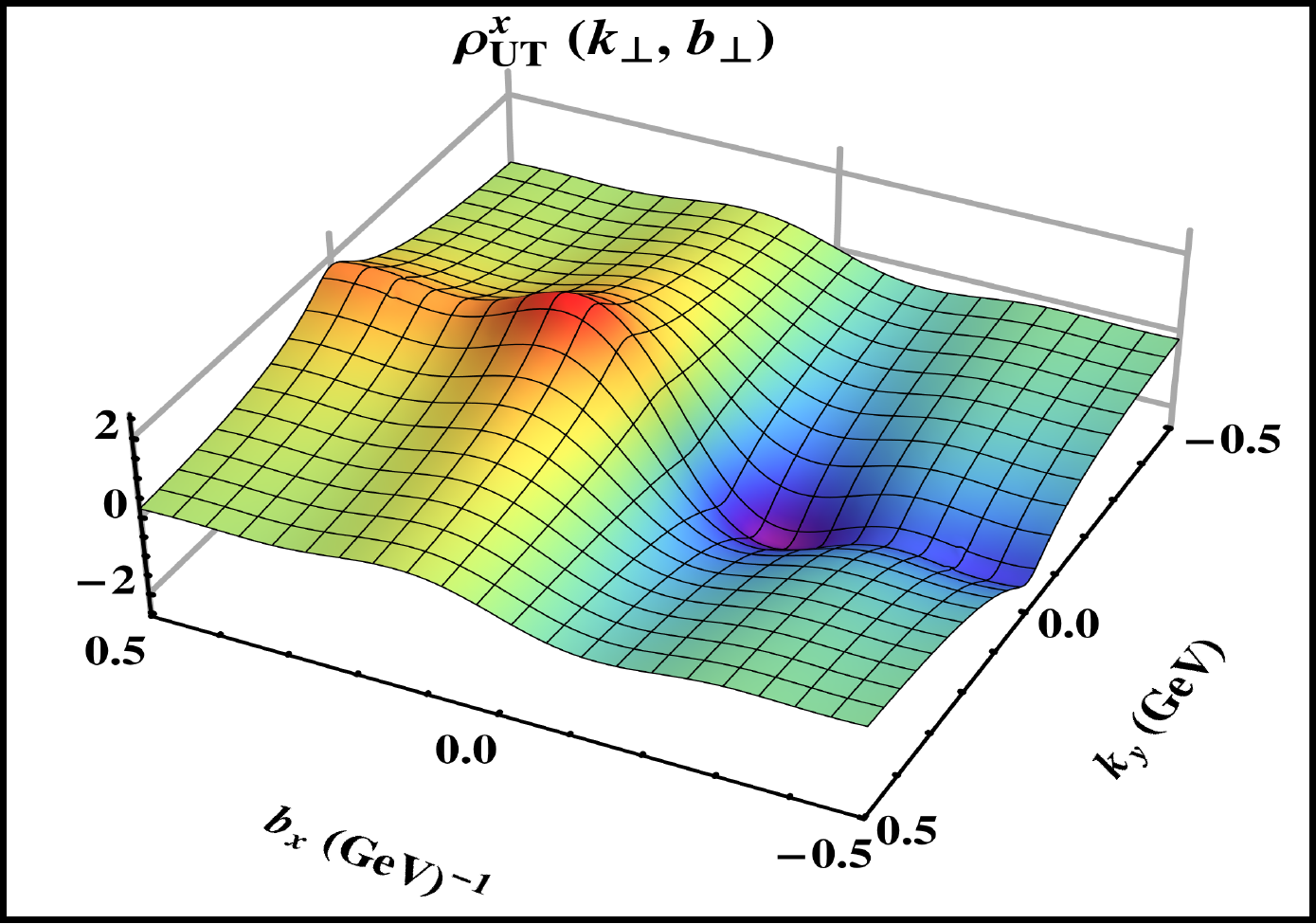}\hskip 0.3cm(f)\includegraphics[width=7.3cm,height=4.7cm]{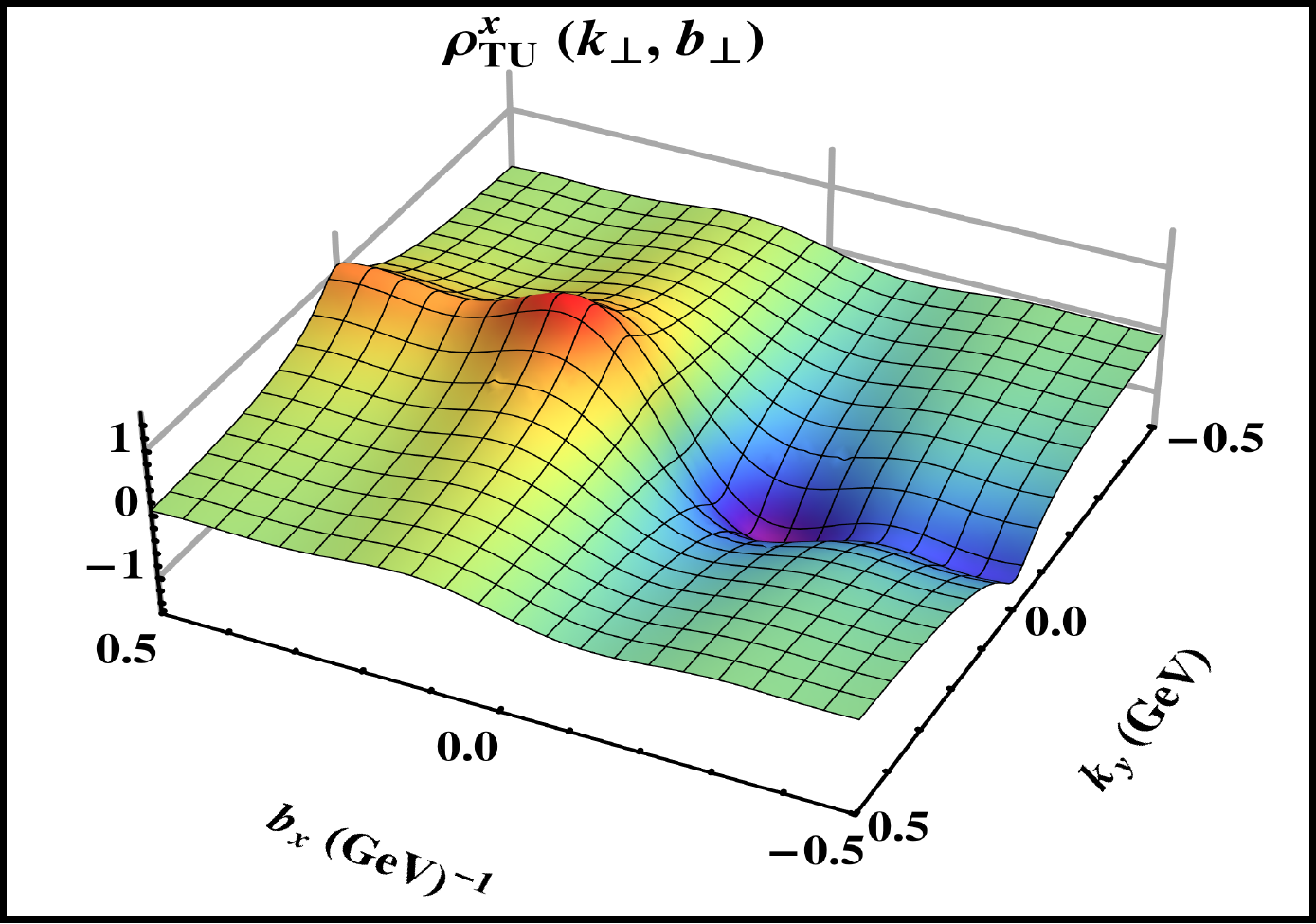}  
  \caption{3D plot of  Wigner distributions $\rho^x_{UT}({\bs k}_\perp, {\bs b}_\perp)$ and 
  $\rho^x_{TU}({\bs k}_\perp, {\bs b}_\perp)$ at $\Delta_{max} = 20~\mathrm{GeV}$. The first row displays the two distributions
  in ${\bs b}_\perp$ space with ${\bs k}_\perp= 0.4~\mathrm{GeV} \,\hat{{\bs e}}_y$. The second row shows the two distributions
  in ${\bs k}_\perp$ space with ${\bs b}_\perp= 0.4~\mathrm{GeV}^{-1}\, \hat{{\bs e}}_y$. The last row represents the two distributions in mixed space.}
 \label{rhout}
\end{figure}
\begin{figure}[h]
 \centering 
(a)  \includegraphics[width=7.3cm,height=4.7cm]{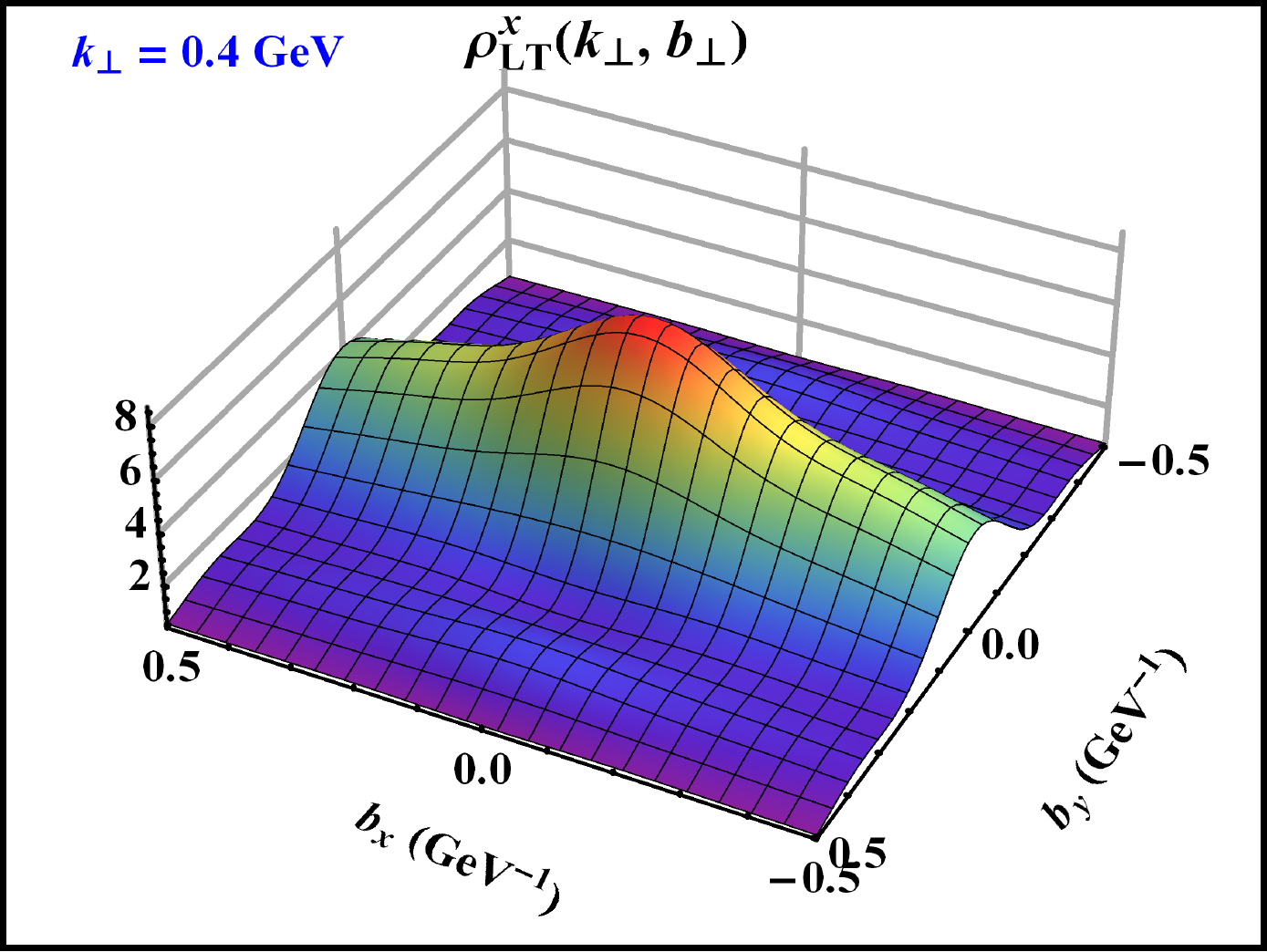}\hskip 0.3cm(d)\includegraphics[width=7.3cm,height=4.7cm]{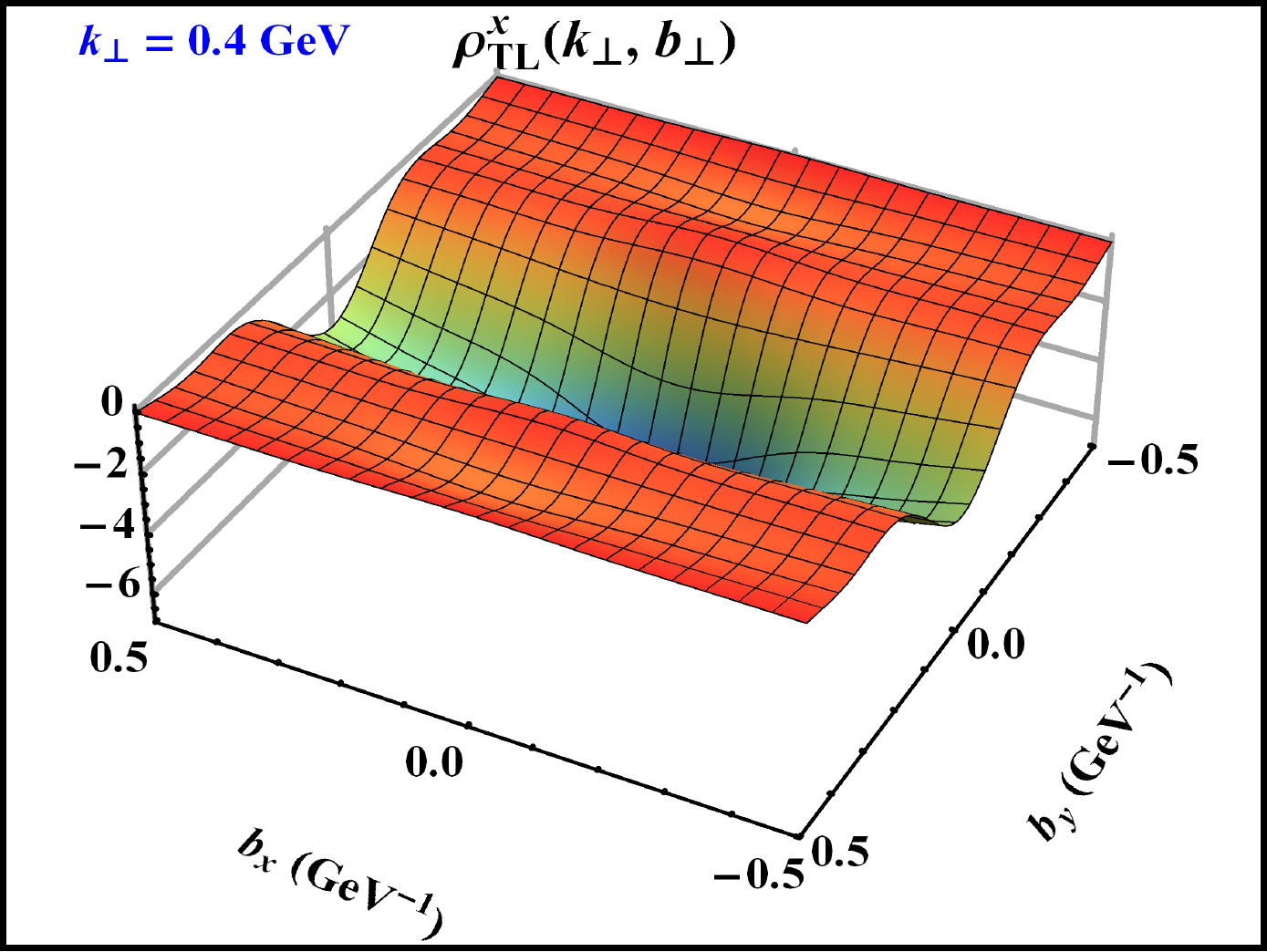}\\[5ex]
(b)\includegraphics[width=7.3cm,height=4.7cm]{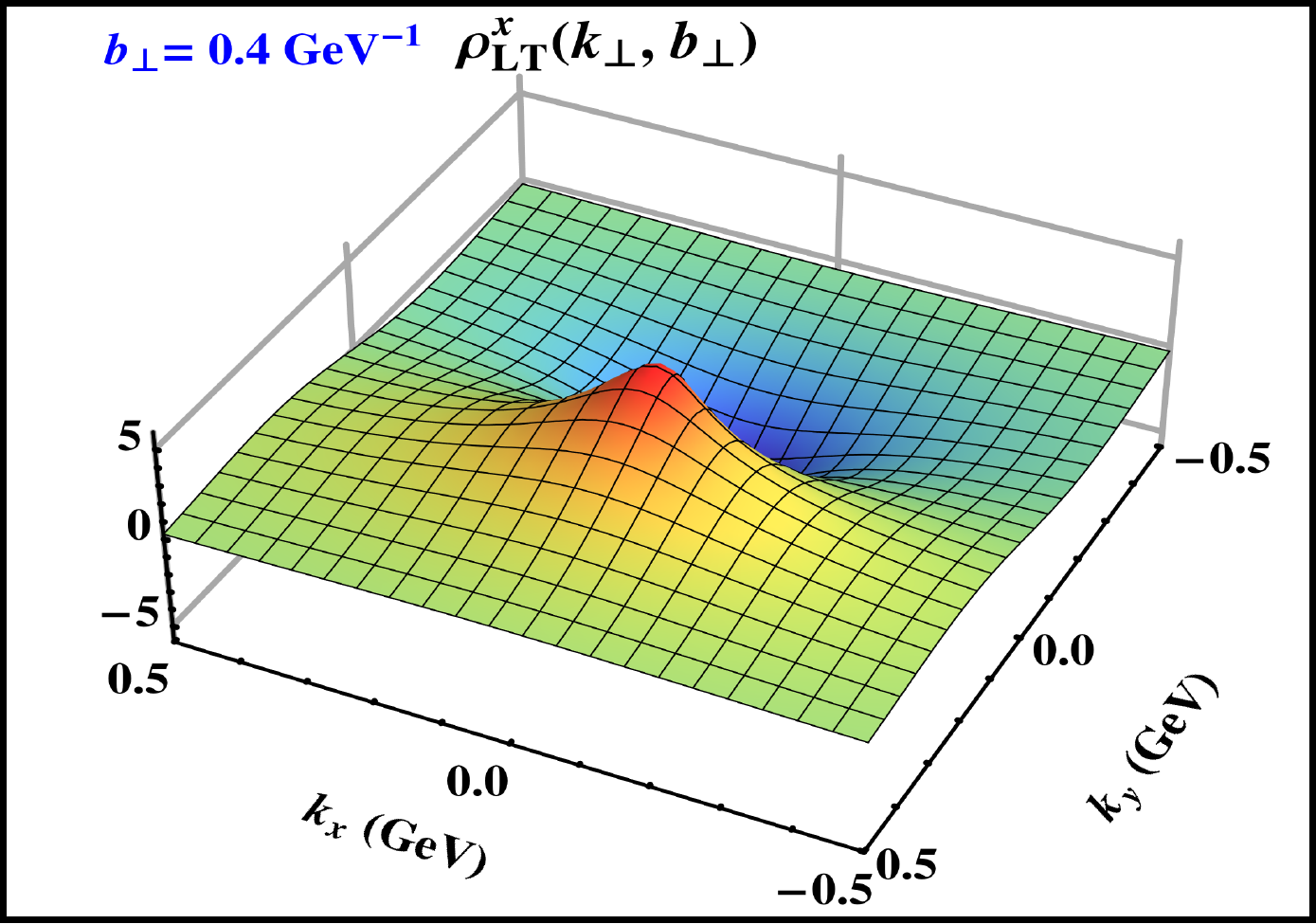}\hskip 0.3cm(e)\includegraphics[width=7.3cm,height=4.7cm]{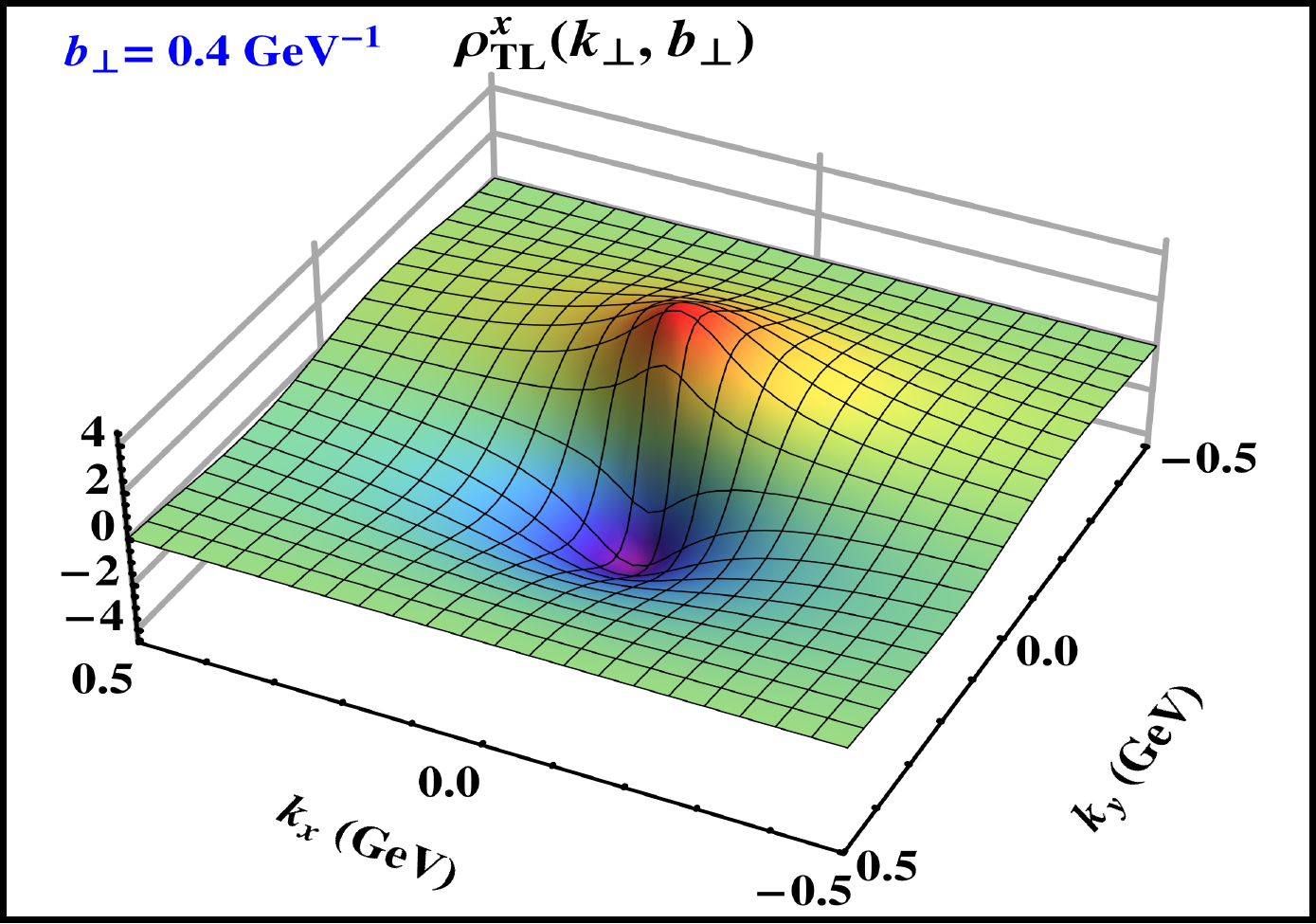}\\[5ex]
(c) \includegraphics[width=7.3cm,height=4.7cm]{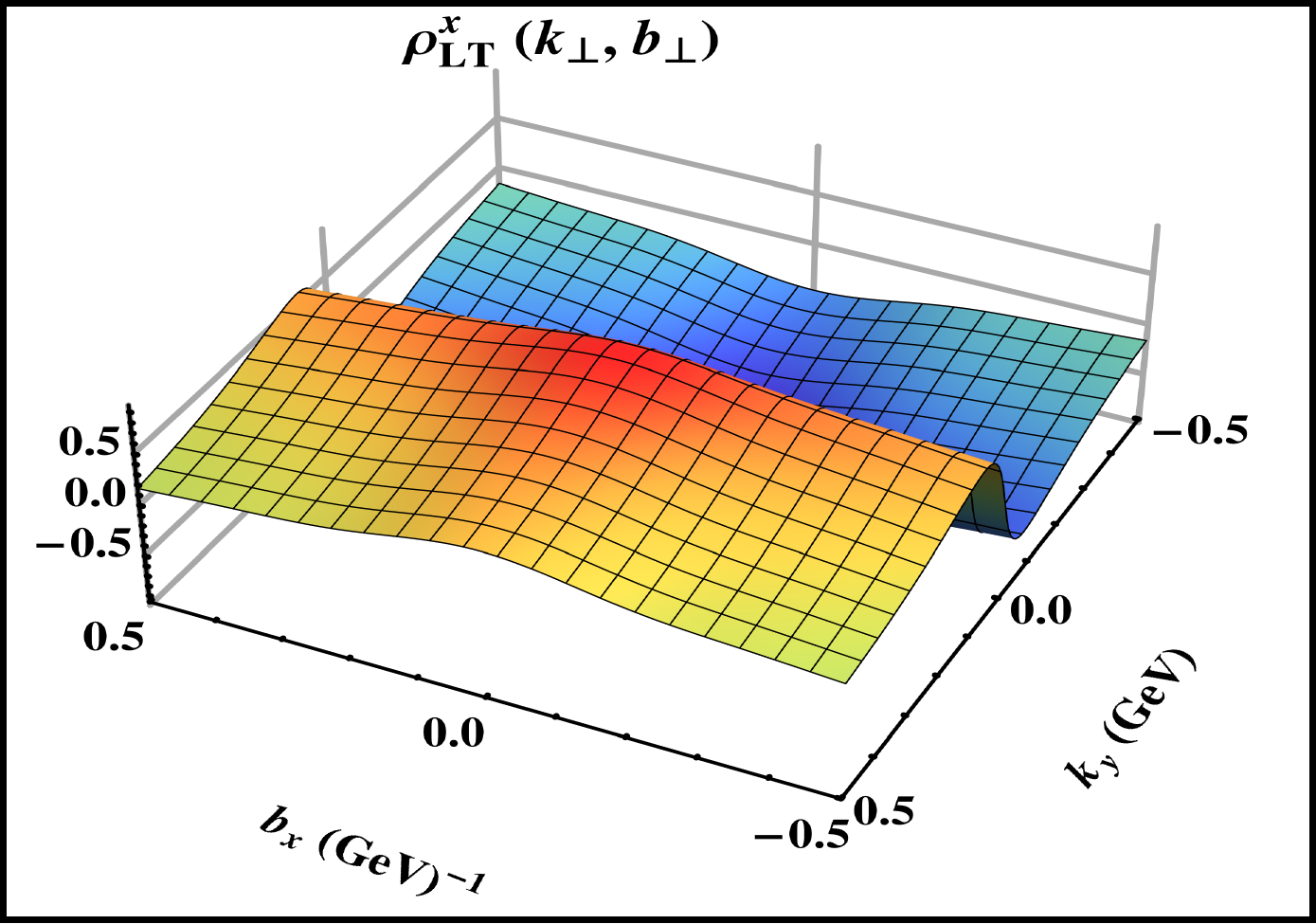}\hskip 0.3cm(f)\includegraphics[width=7.3cm,height=4.7cm]{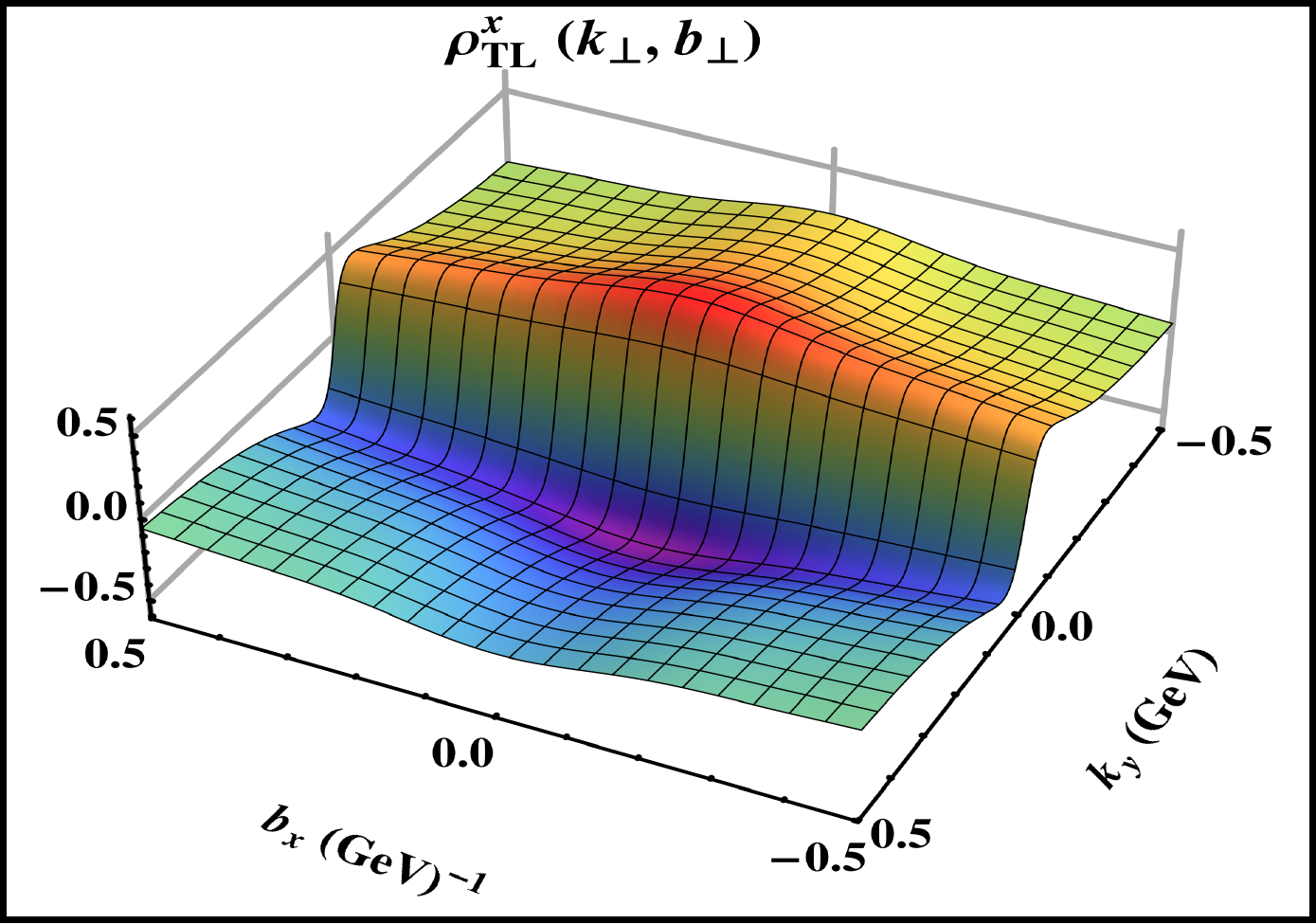}  
  \caption{3D plot of  Wigner distributions $\rho^x_{LT}({\bs k}_\perp, {\bs b}_\perp)$ and 
  $\rho^x_{TL}({\bs k}_\perp, {\bs b}_\perp)$ at $\Delta_{max} = 20~\mathrm{GeV}$. The first row displays the two distributions
  in ${\bs b}_\perp$ space with ${\bs k}_\perp= 0.4~\mathrm{GeV} \,\hat{{\bs e}}_y$. The second row shows the two distributions
  in ${\bs k}_\perp$ space with ${\bs b}_\perp= 0.4~\mathrm{GeV}^{-1}\, \hat{{\bs e}}_y$.
  The last row represents the two distributions in mixed space.} 
 \label{rholt}
\end{figure}
\section{Results and Discussion}\label{result}
In this section, we start by discussing  $\rho_{UU}({\bs k}_\perp, {\bs b}_\perp)$, which is the 
Wigner distribution for an unpolarized quark in an unpolarized dressed quark state.
Figure.~\ref{rhou}(a) shows the distribution in ${\bs b_\perp}$ space with a fixed transverse momentum 
${\bs k_\perp}=0.4~\mathrm{GeV}~\hat{\bs e}_j$. 
We observe a positive peak centered around $b_x=b_y=0$ as 
observed in Refs.\cite{Lorce16, Liu15}.
In ${\bs k_\perp}$ space, we obtain a sharp negative peak shown in Figure.~\ref{rhou}(b).
 Figure.~\ref{rhou}(c) shows the Wigner distribution $\rho_{UU}({\bs k}_\perp, {\bs b}_\perp)$ in mixed space where we have integrated out 
$k_x$ and $b_y$ dependence, thus, we get the probability densities in the $b_x$ - $k_y$ plane. 
The Wigner distribution $\rho_{UU}$ can be related to unpolarized GPD and the unpolarized TMDs by taking the appropriate limit. 

In Figure.~\ref{rhou}(d), we present $\rho_{UL}({\bs k}_\perp, {\bs b}_\perp)$, which is the  Wigner
distribution for a longitudinally polarized quark in an unpolarized dressed quark state in ${\bs k}_\perp$ space with ${\bs b_\perp}= 0.4~\mathrm{GeV}^{-1}~\hat{\bs e}_j$. 
Figure.~\ref{rhou}(e) shows the distribution in the impact parameter space with ${\bs k_\perp}= 0.4~\mathrm{GeV}~\hat{\bs e}_j$.
Figures.~\ref{rhou}(d) and 
\ref{rhou}(e) have a similar nature with opposite polarities. 
These two graphs show a dipole structure, as observed in other models \cite{Lorce11, Liu15}. 
Figure.~\ref{rhou}(f) shows the three-dimensional plot of Wigner distributions
in mixed space which exhibit a quadrupole structure. This distribution can be related to the spin-orbit correlation and orbital angular momentum
of quark as demonstrated in Ref. \cite{Asmita14}, and qualitatively, are in agreement with the chiral quark-soliton model and the constituent quark model. 

In Figures.~\ref{rhol}(a)$-$ \ref{rhol}(c), we plot $\rho_{LL}({\bs k}_\perp, {\bs b}_\perp)$  in ${\bs k}_\perp$ space, ${\bs b}_\perp$ space, and the mixed space,
respectively. Figure \ref{rhol} for $\rho_{LL}({\bs k}_\perp, {\bs b}_\perp)$ shows a similar nature as in Figure~\ref{rhou} for $\rho_{UU}({\bs k}_\perp, {\bs b}_\perp)$ 
since these two distributions only differ in the sign of the mass term whose contribution is negligible compared to other terms.
Hence, numerically we obtain slightly different maxima for them.

 If one integrates the Wigner function $\rho_{UU}$ over ${\bs k}_\perp$ and ${\bs b}_\perp$, one would get the familiar plus distribution ${1\over {(1-x)}}_+$ \cite{Hari99} as expected in the parton distribution of a dressed quark. 
 It is very important to take into account the contribution from the normalization of the state to get the correct behavior at $x=1$. In \cite{Hatta14} the authors considered the two-particle contribution to $\rho_{UU}$ 
 for a dressed quark for fixed $x$, and observed that the negative peak at ${\bs b}_\perp=0$ is due to the fact that for large values of $\Delta_{max}$, the second term in the numerator of the Wigner distribution, which is 
 proportional to ${(\Delta_\perp)}^2$ dominates over the first term. As it comes with a negative sign, there is a large negative peak.  The authors proposed to study the Husimi distributions, which in effect have a Gaussian 
 regularization factor in the integrand that keeps them positive in the entire range of ${\bs b}_\perp$. The Husimi distributions, however, have the limitations that upon integration over ${\bs b}_\perp$ they do not reduce to any known 
 TMD, but upon integration over both ${\bs k}_\perp$ and ${\bs b}_\perp$ they give the parton distributions. Here we see that when integrated over the entire region of $x$, the two-particle sector of the dressed quark model gives a 
 positive peak for the Wigner distribution, similar to other models.
     
Figures~\ref{rhol}(d)$-$ \ref{rhol}(f) show $\rho_{TT}({\bs k}_\perp, {\bs b}_\perp)$, which describes the distribution when the quark and dressed quark
state both are  transversely polarized. In this case, we can have two independent distributions. One is when both the quark
and the dressed quark are polarized parallelly in, say, the $x$-direction. The other is the pretzelous Wigner distribution
when the quark and dressed quark are transversely polarized along the two orthogonal directions. In our model, the 
latter distribution is zero. So we study only the former case in ${\bs k}_\perp$ space, 
${\bs b}_\perp$ space, and mixed space, respectively. 
It is important to note that the nature of $\rho_{TT}({\bs k}_\perp, {\bs b}_\perp)$ is similar to 
$\rho_{UU}({\bs k}_\perp, {\bs b}_\perp)$, and $\rho_{LL}({\bs k}_\perp, {\bs b}_\perp)$ and this can be
inferred from the analytical expressions Equations (\ref{rhouu}), (\ref{rholl}), and ~(\ref{rhott}) . 
Behavior of $\rho_{TT}$  is similar to the one obtained in Ref. \cite{Liu15}, which was calculated in a spectator model. 

Figures~\ref{rhout}(a)$-$  \ref{rhout}(c) show the three-dimensional plot of the Wigner distribution $\rho^x_{UT}({\bs k}_\perp, {\bs b}_\perp)$ in ${\bs k}_\perp$ space, 
${\bs b}_\perp$ space, and mixed space, respectively. These distributions account for the transversely polarized quark in an unpolarized
target state and the quark polarization is taken as the $x$-direction. In the TMD limit, we observe that the $\rho^x_{UT}$ distribution vanish, as expected in our model, as we have not taken into account the gauge link, and so cannot get the T-odd distributions. 
Figures~\ref{rhout}(d)$-$  \ref{rhout}(f) show the three-dimensional plot of Wigner distribution $\rho^x_{TU}({\bs k}_\perp, {\bs b}_\perp)$ in ${\bs k}_\perp$ space, 
${\bs b}_\perp$ space, and mixed space, respectively. These distributions describe the unpolarized quark in a transversely polarized
target state and the referred direction in the transverse plane is the $x-$direction here.
We observe that $\rho^x_{TU}({\bs k}_\perp, {\bs b}_\perp)$ and $\rho^x_{UT}({\bs k}_\perp, {\bs b}_\perp)$ behave
identically in the ${\bs k}_\perp$ space, 
${\bs b}_\perp$ space, and mixed space. The functional dependence of these two distributions only differ by a 
factor of the $x$ in the numerator, and the contribution coming from this $x$ dependence is not that significant compared to the $\Delta_{\perp}$ term which dictates the overall nature of the plot. 
In ${\bs b}_\perp$ space, for $\rho^x_{TU}({\bs k}_\perp, {\bs b}_\perp)$ and $\rho^x_{UT}({\bs k}_\perp, {\bs b}_\perp)$ we observe a dipole nature and since the ${\bs b}_{\perp}$ dependence is entirely contained inside the $\mathrm{sine}$ factor, the sign flip required for the dipole behavior is governed by the property of the $\mathrm{sine}$ function. 
In ${\bs k}_\perp$ space we see a quadrupole nature in the 3D plots. The ${\bs k}_{\perp}$ dependence for $\rho^x_{TU}$ and $\rho^x_{UT}$ is confined within the denominator term denoted by $D({\bs q}_{\perp})$ and $D({\bs q}'_{\perp})$ which
means the quadrupole behavior is due to the dot product ${\bs k}_{\perp}\cdot{\bs \Delta}_{\perp}$ residing in those terms.
In mixed space, we find that both $\rho^x_{TU}$ and $\rho^x_{UT}$ show a dipolelike behavior. We also note that these distributions in 
${\bs b}_\perp$ space behave similar to the spectator model results in \cite{Liu15} and behave differently in 
${\bs k_\perp}$ space and in mixed space.  \\
\indent Finally, Figures~\ref{rholt}(a)$-$ \ref{rholt}(c) describe the transverse Wigner distribution $\rho^x_{LT}({\bs k}_\perp, {\bs b}_\perp)$ in ${\bs k}_\perp$ space, 
${\bs b}_\perp$ space, and mixed space, respectively. These distributions describe a transversely polarized quark in a longitudinally
polarized target state and here the direction of the polarization of the quark is referred in the $x$-direction. 
Figures~\ref{rholt}(d)$-$ \ref{rholt}(f) describe the transverse Wigner distribution $\rho^x_{TL}({\bs k}_\perp, {\bs b}_\perp)$ in ${\bs k}_\perp$ space, 
${\bs b}_\perp$ space, and mixed space, respectively. These distributions describe a longitudinally polarized quark in a transversely
polarized target state and the target is polarized in the $x$-direction. 
As was the case with $\rho^x_{TU}({\bs k}_\perp, {\bs b}_\perp)$ and $\rho^x_{UT}({\bs k}_\perp, {\bs b}_\perp)$, 
$\rho^x_{LT}({\bs k}_\perp, {\bs b}_\perp)$ and $\rho^x_{TL}({\bs k}_\perp, {\bs b}_\perp)$ only differ by a factor of $x$ and additionally they have a sign difference which is reflected in all the 3D plots.
Again the contribution coming from this difference in $x$ dependence is not significant enough to show up in the 3D plots.
In ${\bs b}_\perp$ space we observe the expected behavior modulated by the $\mathrm{cosine}$ term. There is maximum at $b_x=b_y=0$ for $\rho^x_{LT}({\bs k}_\perp, {\bs b}_\perp)$ which gets flipped into a minimum for $\rho^x_{TL}({\bs k}_\perp, {\bs b}_\perp)$.
Both in ${\bs k}_\perp$ space and mixed space we observe a dipolelike behavior, but the mixed space dipole is more spread out compared to the ${\bs k}_\perp$ space. 
These two  distributions behave in the same way as the spectator model \cite{Liu15} in ${\bs k_\perp}$ space and differently in ${\bs b}_\perp$ space. 
Again, such results depend on model parameters. \\
In our model, we do not consider the multipole decomposition as discussed in 
Ref. \cite{Lorce16}, which is a model-independent way of studying the Wigner distribution. 
The behavior of $\rho^x_{UT}$, $\rho^x_{TU}$, $\rho^x_{LT}$, $\rho^x_{TL}$, we obtain, 
concurs with \cite{Lorce16}. For example, we studied the terms of the type 
$\rho^{(0,1)}_{LTx} \,\,\alpha \,\,k_x$ and found them in good agreement with our model. 
It would be interesting to compare the multipole decomposition in our model but in this 
work we have limited our discussion to terms proportional to either ${\bs k}_\perp$ or 
${\bs \Delta}_\perp$ (which is the conjugate to ${\bs b}_\perp$); hence we do not obtain the dipole 
and quadrupole behavior as observed  by the model-independent analysis of all 
these distributions in Ref. \cite{Lorce16}.
\section{Conclusion}\label{conclusion}  
In this work, we include transverse polarization of the target state and quark to calculate the Wigner distributions, 
unlike in the previous work \cite{Asmita14}. Thus, we have now 
studied Wigner distributions of quark in different polarizations in the dressed quark model using LFWFs. 
We have used an improved method for the numerical integration that gives better convergence of the results, and the dependence on   $\Delta_{max}$ present in our earlier work \cite{Asmita14, Asmita15} is removed.  
Wigner distributions contain information that one cannot extract from GPDs and TMDs, as they may contain a  correlation between 
quarks and gluons in transverse position and  three-momentum. As they have not been accessed in experiments yet,  model-based calculations are important to gain insight into them. Equivalently, as both the GPDs and TMDs are linked to Wigner distributions
and there are experimental data available on observables dependent on the GPDs and TMDs,  these connections can help us in formulating
better phenomenological models that are closer to reality, thereby giving us a better understanding
of hadron physics. 

We calculate twist-2 quark Wigner distributions for a quark state dressed at one loop by a gluon, which can be thought of as a field theory based model of a composite relativistic spin-1/2  state. 
 We have considered unpolarized, longitudinal and transverse polarization combinations for both the quark and the target state.
 We obtain eight independent quark Wigner distributions in our model.  The pretzelous distribution $\rho_{TT}^{\perp}(x,{\bs k}_\perp, {\bs b}_\perp)$ was found to vanish in our model. 
 The unpolarized $\rho_{UU}({\bs k}_\perp, {\bs b}_\perp)$, longitudinally polarized $\rho_{LL}({\bs k}_\perp, {\bs b}_\perp)$,
and the tranversity distributions $\rho_{TT}({\bs k}_\perp, {\bs b}_\perp)$ show a similar nature. One can obtain the unpolarized GPD and TMDs from 
an unpolarized distribution. We found in this model that  $\rho_{LU}({\bs k}_\perp, {\bs b}_\perp)$ is equal to  $\rho_{UL}({\bs k}_\perp, {\bs b}_\perp)$.
These distributions can be related to the spin-orbit correlation and orbital angular momentum of quark, as discussed in Ref. \cite{Asmita14}. We also observed that $\rho^x_{TU}({\bs k}_\perp, {\bs b}_\perp)$ and $\rho^x_{UT}({\bs k}_\perp, {\bs b}_\perp)$ exhibit a similar nature as they differ 
only by a factor $x$ and can be seen from 3D plots. Similarly, $\rho^x_{TL}({\bs k}_\perp, {\bs b}_\perp)$ and $\rho^x_{LT}({\bs k}_\perp, {\bs b}_\perp)$
also differ by a factor $x$ with the signs flipped, which can be seen from the 3D plots. 
In some cases, our results in this perturbative model differ qualitatively from a previous calculation in the spectator model. 
In some Wigner distributions and mixed distributions, we perceived dipole and quadrupole structures. 
Further work in the model would be to calculate the gluon Wigner distributions, with all possible polarization configurations at the leading twist. \\
{\bf Acknowledgements} \\
We would like to thank Oleg Teryaev for fruitful discussions. 

\end{document}